\documentclass[12pt]{article}
\usepackage{biblatex} 
\RequirePackage{amsthm,amsmath,amsfonts,amssymb, bm, graphicx, xurl, algorithm}
\usepackage{algpseudocode}
\RequirePackage[colorlinks,citecolor=blue,urlcolor=blue]{hyperref}

\newcommand{\Sn}{\mathcal{S}_n}

\newtheorem{theorem}{Theorem}[section]
\newtheorem{assumption}{Assumption}[section]
\title{Local Variable and Neighborhood Selection for Firearm Fatality in South-East USA}
\usepackage{hyperref}
\author{
Debjoy Thakur\thanks{Department of Statistics and Data Science, Washington University in St. Louis. Email: \texttt{debjoy@wustl.edu}. ORCID: 0000-0001-5514-5707}
\and
Lingyuan Zhao\thanks{Department of Statistics and Data Science, Washington University in St. Louis. Email: \texttt{lingyuan@wustl.edu}.}
\and
Soutir Bandyopadhyay\thanks{Department of Applied Mathematics and Statistics, Colorado School of Mines. Email: \texttt{sbandyopadhyay@mines.edu}. ORCID: 0000-0003-2213-3333. \textbf{Corresponding author.}}
}

\date{} 

\begin{document}

\maketitle

\begin{abstract}
A major public health concern in the United States (US) is gun-related deaths. The number of gun injuries largely varies spatially because of county-wise heterogeneity of race, sex, age, and income distributions. But still, a major challenge is to locally identify the influential socio-economic factors behind these firearm fatality incidents. For a diverging number of predictors, a rich literature exists regarding SCAD under the independence framework; however, a vacuum remains when discussing local variable selection for spatially correlated, over-dispersed data. This research presents a two-step localized variable selection and inference framework for spatially indexed gunshot fatality data. In the first step, we select variables locally using the SCAD penalty for specific locations where the number of gunshot incidents exceeds a threshold. For these locations, after selecting the predictors, we proceed to the next step, which involves examining the directional variation in the latent spatial neighborhood structure. We further discuss the theoretical properties of this county-specific local variable selection under infill asymptotics. This method has threefold advantages: (i) this method selects the variables locally, (ii) this method provides inference about directional variation of a selected predictor, and (iii) instead of assuming the spatial neighborhood structure in an ad hoc manner, this method identifies the specific type of spatial neighborhood structure that is most appropriate for modeling the random effects.
\end{abstract}


\section{Introduction}
In the US, a serious concern in public health is gun violence-related early death. This gun-related death is generally stored as a count of data per 100,000 people. Many public health reports have reported that in the year 2022, there were almost 54\% gun-related deaths due to suicides and 43\% due to homicides. These gun-related death numbers are significantly higher in the US compared to other First World countries. In gun-related deaths, the heterogeneity has arisen mainly because of the heterogeneity in race, age, and sex distributions in the US. According to \cite{villarreal2024gun} in 2022, a significant rise in black young people belonging to the age group of 15 to 34 years old, which is nearly one-third of the total gun-related deaths in the US, although the total Black population only consists of $2\%$ of the total population, and there is a recent rise in gun-suicidal deaths in the white male population whose age is in between 55 and 74 years. Their murder rate was nearly twenty-four times greater than that of white men in the same age group. Black teenage girls and young women were around nine times more likely than white teenage girls and young women to be killed by a gun. In the last ten years, gun murders among Hispanic/Latino men have risen significantly from 2013 to 2022, by about 70\%. Among American Indians and Alaska Natives, the gun-related murders were five times higher than those of white people. In the context of gun-related suicides \cite{villarreal2024gun} has reported that among Hispanics and Latinos, gun-related suicide has increased by more than $50\%$ since 2013, and this suicidal tendency among white people is higher, which results the $70\%$ of the total deaths. This discussion briefly outlines how the variation of gun deaths is affected by the variation of age, race, and sex distribution.

There is a lot of research on penalized methods for choosing variables and models for independent observations. Some examples are LASSO \cite{lassotbshirani}, smoothly clipped absolute deviation (SCAD) \cite{Fan01122001}, the Dantzig selector \cite{candes2007dantzig}, and the minimax concave penalty (MCP) \cite{mcp}. The variable selection consistency of LASSO has been thoroughly investigated; \cite{knight2000asymptotics}, \cite{zhao2006model}, and \cite{lahiri-vsc} have provided substantial theoretical advancements. The works of \cite{hoeting2006model} and \cite{huang2007optimal} are the earliest articles that discuss the problem of variable selection for spatial data. The first one employed Akaike's information criterion (AIC) to choose covariates for geostatistical regression models in the presence of spatial correlation. The second suggestion was to use generalized degrees of freedom (GDF) to choose spatial models. To simultaneously choose predictors and neighborhood structures within a defined geographic dependency, \cite{huang2010spatial} subsequently developed a spatial extension of LASSO, despite the absence of corresponding theoretical guarantees. By examining the asymptotic characteristics of penalized least squares estimators utilizing several penalty functions, \cite{wang2009variable} advanced theoretical inquiries into penalized estimation within spatial contexts. An adaptive LASSO framework for variable selection in spatial linear models for Lattice data, under the Gaussianity assumption, was established by \cite{zhu2010selection}. For spatial linear models, \cite{pmle} developed penalized maximum likelihood estimation with tapered spatial covariance functions as a likelihood-based alternative. The methodology of \cite{zhu2010selection} was extended by \cite{reyes2012selection} in the context of spatio-temporal lattice data. \cite{fu2013estimation} introduced a penalized pseudolikelihood methodology for variable selection in autologistic regression models, focusing on discrete spatial outcomes. Recent advancements include GMM-based LASSO estimators for autoregressive spatial error models \cite{cai2019variable} and generalized LASSO-type techniques for spatial additive models employing group and adaptive group penalties \cite{spatial-additive}. \cite{zhao2020solution} advanced algorithms for generalized LASSO problems by reformulating them to be efficiently solvable using conventional LASSO solvers such as least angle regression (LAR) and glmnet, eliminating the necessity for full-rank penalty matrices. \cite{li2020variable} examined the selection of variables for spatial semiparametric models with SCAD penalties, while \cite{cai2020variable} investigated the process of drawing inferences after penalized estimates in geographical models. Recently, spatial variable selection has been broadened to include multivariate and complex dependent systems. Notably, \cite{maranzano2023adaptive} used adaptive LASSO for functional hidden dynamic geostatistical models within mixed-effects frameworks, and \cite{desai2024covariance} presented a covariance-assisted multivariate penalized additive model that supports simultaneous variable selection and inference. Penalized techniques for spatial error models \cite{al2017penalty}, conditional autoregressive (CAR) models \cite{gonella2022facing}, and spatiotemporal autoregressive models \cite{safikhani2020spatio} constitute further contributions.

A common limitation of this earlier study is that the researchers did not sufficiently consider local variable selection or post-selection inference for the spatial random-effects structure. To the best of our knowledge, this is the first work to perform \textit{local variable selection for overdispersed areal data}. Further, as an extension, we identify an appropriate spatial neighborhood structure that incorporates directional variation. In particular, we make the following contributions:

\begin{enumerate}
\item We introduce a two-stage approach. In the first stage, we use the SCAD penalty to select predictors locally, assuming no random effects.
\item In the second stage, we estimate the regression coefficients and the neighborhood topology using the rescaled, selected predictors.
\item Beyond local variable selection, we also characterize directional variation in the coefficient surface, which helps identify the directions along which a selected predictor exerts greater influence.
\end{enumerate}
%
The remainder of the paper is organized as follows. In Section~\ref{s:motivation}, we discuss the motivation for studying local variable selection in the context of gun-related deaths. Section~\ref{s:methods} provides a detailed description of our methodological contributions, along with computational details for implementing local SCAD. In Section~\ref{s:theory}, we establish theoretical properties of the local SCAD-based estimator under infill asymptotics. Sections~\ref{s:simulation} and~\ref{s:real-data} present results from the simulation study and the real-data analysis, respectively. Finally, in Section~\ref{s:conclusion}, we conclude with a discussion of key insights and future directions.

\section{Motivation}\label{s:motivation}
As per the report of \cite{villarreal2024gun} in the year 2022, gun violence killed over 48,000 people in the United States, about 132 deaths a day, or one death every eleven minutes. Since 2013, there has been a significant increase in gun-related deaths, which has been mentioned in many reports, and more specifically, the number of teenage deaths due to guns has doubled. According to \cite{davis2023us}, the Centers for Disease Control and Prevention (CDC) used the CDC WONDER record and arranged the death records and their reasons altogether in their report. The final report related to national gun-related mortality data of 2022 was made publicly available in April 2024.

Geographic dynamicity makes the spatial patterns of gun deaths more complicated, and a clear example of this situation is where Mississippi had the highest rates, while Massachusetts had the lowest rate of firearm deaths in 2022. In most states, gun homicides were the prime reason for juvenile gun deaths, whereas in some places, like Idaho, the scenario is different; here, the gun suicide is mainly responsible, as mentioned in \cite{villarreal2024gun}. There were also variations between rural and urban areas: rural counties had the highest rates of firearm suicides, whereas major metropolitan counties had the highest rates of firearm homicides. These patterns reveal how socioeconomic factors, such as demographics, gun access, and policy contexts, work together. 
\begin{figure}[h!]
        \centering
        \includegraphics[width=\linewidth]{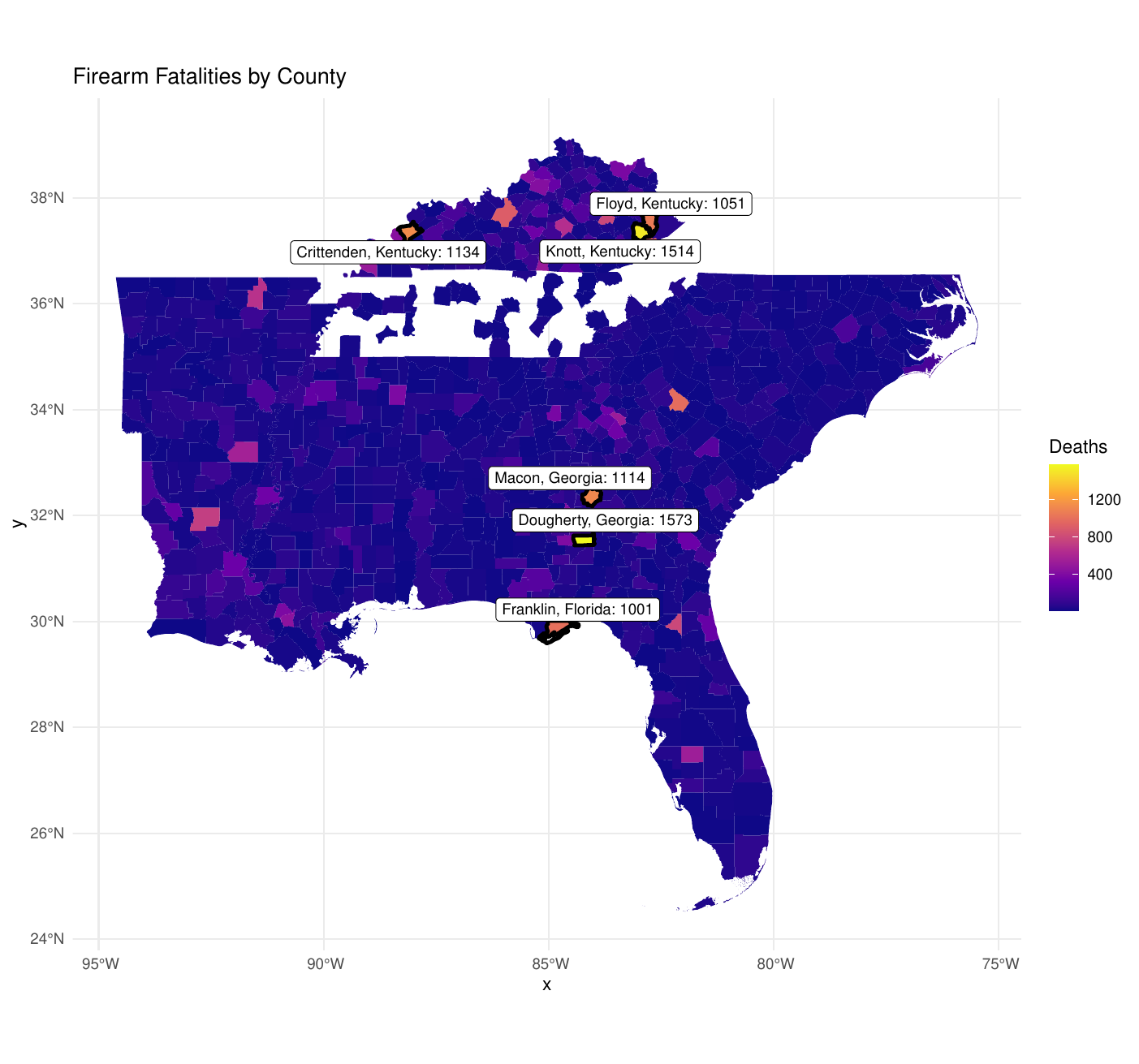}
        \caption{County-wise Firearm fatalities in 2024 in the Southeast USA.}
        \label{fig:firearm_2024}
\end{figure}
In this research, the firearm fatalities in the Southeast USA in the year 2024 have attracted our attention mostly. From this accessible report \url{https://everytownresearch.org/report/city-data/} and the references mentioned in there, we observe that about 44,400 individuals died in gun-related events in 2024. Although there is good news that this gun-related death rate has decreased by 4\% since 2023. The main reason for this drop is the lower number of murders and mass shootings. In certain big cities, the number of murders and mass shootings has declined by 20\% and 24\% since 2023, but still, the number of gun suicides will rise for the sixth consecutive year. According to the report at \url{https://counciloncj.org/crime-trends-in-u-s-cities-mid-year-2025-update/}, under the presidency of President Joe Biden, the federal government prioritized many programs to reduce gun violence. As a result, gun deaths were reduced in 2023 and 2024. Among many initiatives, there were two remarkable steps: one is the creation of a White House Office of Gun Violence Prevention department to take tougher actions against gun dealers who fail to comply with regulations, and the second one is the enactment of the 2022 Bipartisan Safer Communities Act, the most significant federal gun law in decades. This law helps in gathering financial support to stop violence in communities, makes background checks for younger buyers stronger, and includes new ways to stop gun trafficking. At the same time, gun suicides kept going up, reaching about 27,600 deaths in 2024 and making up almost two-thirds of all gun deaths. More than half of all suicides involved guns, which is the greatest percentage seen in at least 25 years. There were also variations between states; for example, Wyoming, Alaska, and Montana had the greatest rates of gun suicides, while Washington, D.C., Hawaii, and New Jersey had the lowest rates. Most of the U.S. cities reported that gun homicides were still going down, and approximately more than 60\% of cities have recovered to the same number of murders as before the pandemic. In 2024, all ten of the biggest cities saw fewer murders than the year before, saving hundreds of lives. Black and Latina people, as well as people living in big cities and the Northeast portion of the country, had the biggest drops. These trends show that there has been real improvement and that there are still problems that need to be solved when it comes to gun violence in different ethnic groups and areas.

\begin{figure}[h!]
        \centering
        \includegraphics[width=\linewidth]{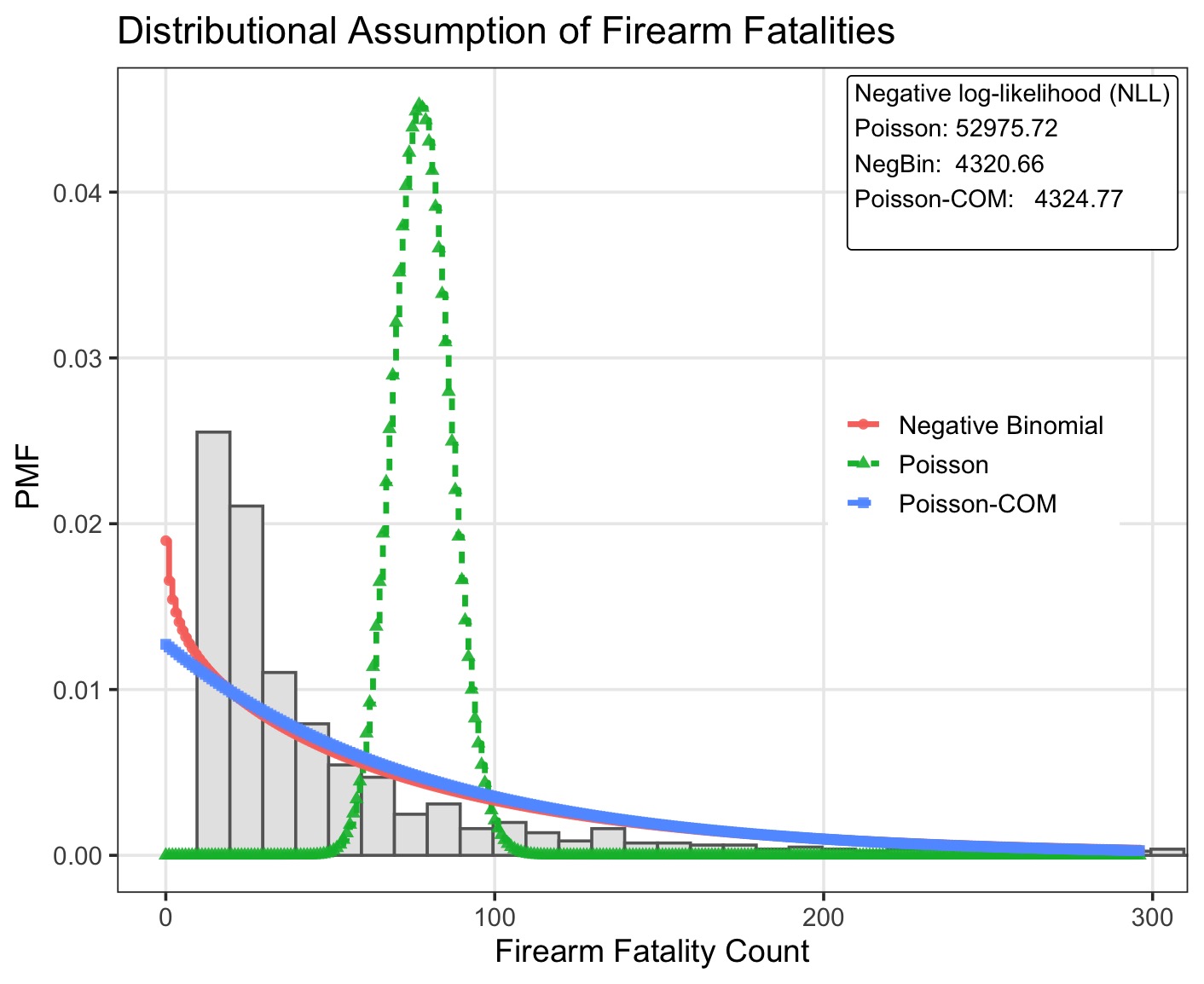}
        \caption{Distributional assumptions for the firearm fatality count.}
        \label{fig:dist_assumption}
 \end{figure}

 We merge census demographic data to analyze the impact of geographic demographic diversity on firearm fatalities for the year 2024. For this research, the county-based demographic data is available in \url{https://www.census.gov/data} for the year 2024. This data contains demographic information on age and sex levels, as well as gender-wise racial distribution in each county. The firearm fatality data is available in \url{https://www.countyhealthrankings.org/health-data/community-conditions/social-and-economic-factors/safety-and-social-support/firearm-fatalities}. These are open-source, accessible data. County-wise firearm fatality data for the Southeast USA is visualized in Figure~\ref{fig:firearm_2024}. In Figure~\ref{fig:firearm_2024}, we observe that in Kentucky, Floyd, Crittenden, and Knott; in Georgia, Macon and Dougherty; and in Florida, Franklin have exceeded the firearm fatality count, which exceeds one thousand firearm fatality incidents in 2024.
In Figure~\ref{fig:dist_assumption}, we see that firearm fatality data is over-dispersed. But for all these counties, not all demographic factors may be equally important. Similarly, while we are talking about spatial random effect, it is also not possible that for each of the hotspot counties, the spatial random effect will be exactly similar, for example, first-order neighborhood structure or second-order neighborhood structure, and so on. 
\begin{figure}[h!]
    \centering
        \centering
        \includegraphics[width=\linewidth]{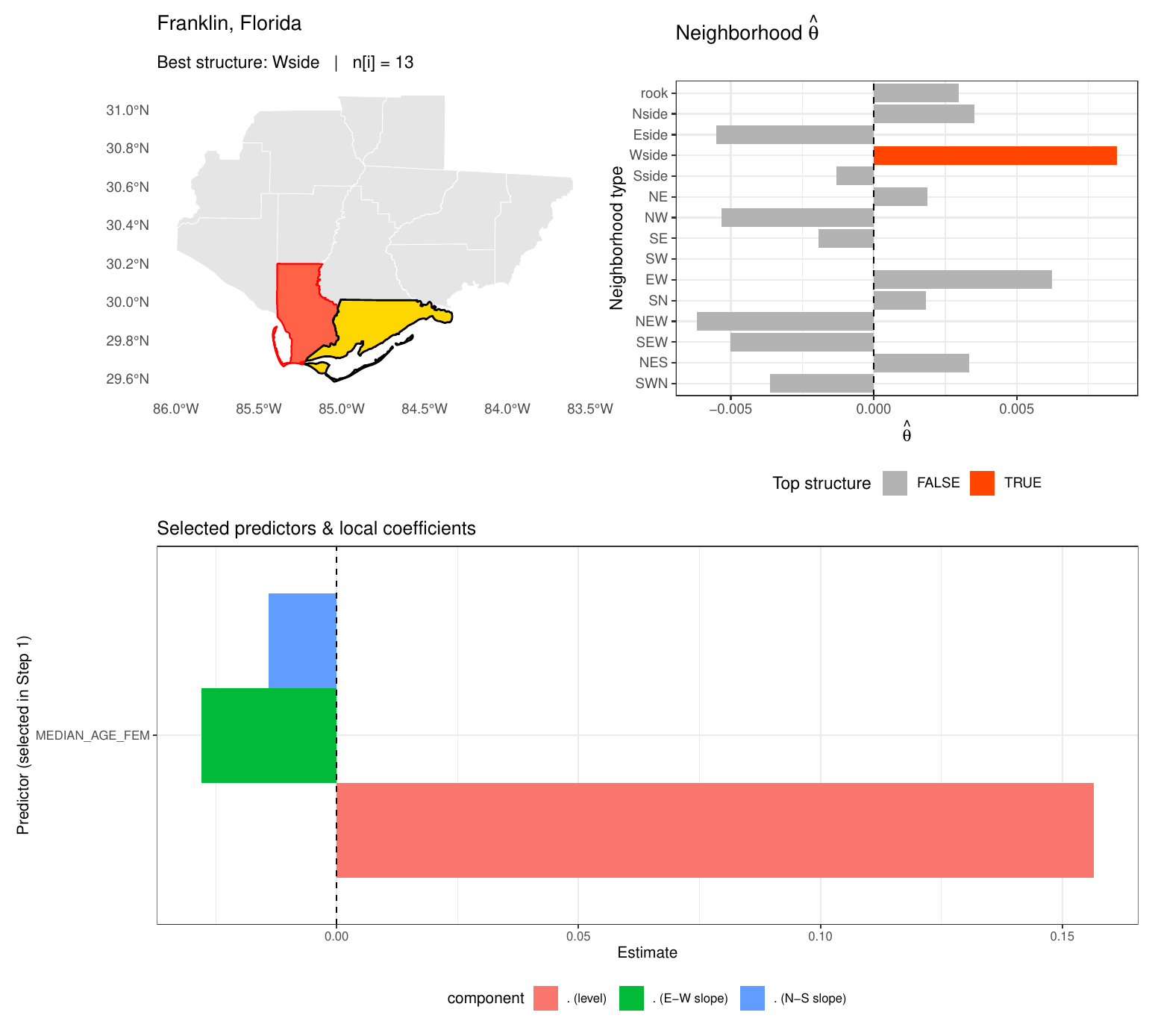}
         \caption{Local neighborhood structure and selected covariates for Franklin County in Florida}
        \label{fig:franklin-florida}
\end{figure}
According to this accessible report about annual gun death data of Florida State, \url{https://usafacts.org/answers/how-many-people-die-from-gun-related-injuries-in-the-us-each-month/state/florida/}, we detect that in 2024, over 3.2 thousand individuals died from gun-related injuries, and there were approximately $13.2$ deaths per $100,000$ person. From Figure~\ref{fig:franklin-florida}, in Franklin County, we see that the chosen covariate is the median age of females, and that predictor is more important along the horizontal direction compared to the vertical direction. We have also observed that instead of a first-order spatial neighborhood for Franklin County in Florida, the spatial random effect is distributed mainly along the west direction. From this detailed discussion, we have formulated three important research problems and will discuss their relevance in the context of public health:
\begin{enumerate}
    \item The socio-demographic diversity always has a significant impact on firearm fatality. In two counties, a variation in socio-demographic features is evident. For example, in one county, a larger number of young Black males die due to homicide, while within the same state, in another county, older white female people die by gun suicide. As a result, while we are focusing on firearm fatality, this diversity of age and race distribution will create different impacts for two different counties. During variable selection in one state, young Black males should be selected, whereas in the other state, the older white female population should be selected. The traditional variable selection method will not be effective in this scenario because they select variables globally. So the question arises, how can we select variables locally?
     \item Now, while we select variables, we do not emphasize the hidden directional variation inherited in the selected coefficient surface. For example, in one county in one direction, say in the east direction, there is a hill station, and within the same county in the other direction, say in the north, there is a populated area. Now, if one variable, say the young male population, is selected for firearm fatality in that county, then the predictor should have a higher regression coefficient along the vertical direction. From this scenario, the second question arises: how can we import the directional variation into the coefficient surface for selected variables? That will provide the information on directional variation in the selected coefficient surface.
   
    \item In general, in spatial statistics, it is quite common in areal data for a spatial generalized linear model to assume about the spatial random effect before initiating the experiment, for example, first order (where four neighboring counties), second order (where eight neighboring counties) neighborhood structures are considered. But this assumed neighborhood structure may not always be justifiable, and there is a possibility of the presence of directional variation in the spatial neighborhood structure. From this scenario, the third question arises: how can we select the hidden neighborhood structure locally? As a result, for two different hotspot counties, there may exist two different types of spatial neighborhood structure.
\end{enumerate}
In the next section, we describe how we solve these three research problems for over-dispersed spatial count data.
\section{Methods}\label{s:methods}
Let's assume that the spatial region $\Sn$ is partitioned into $n$ areal units. In this scenario, $n$ counties are considered as $n$ areal units. Suppose $Y(\bm s)$ is conditionally following a negative binomial (NB) distribution with mean function, $\mu(\bm s):\mathcal{S}_n \to \mathbb{R}$ and positive overdispersion parameter $\sigma$ such that 
\begin{equation}\label{eq:simulated-resp}
    Y(\bm s_i)| \left\{\bm X(\bm s_i), Y(\mathcal N_i)\right\} \sim \mathrm{NB}(\mu_i(\bm X(\bm s_i), Y(\mathcal N_i)), \sigma); \ \bm s_i \in \mathcal{S}_n,
\end{equation}
where we assume the mean function $\mu(\cdot)$ depends on the covariate space, $\bm X = (X_1,\dots, X_{P_n})^\top \in \mathbb{R}^{P_n}$ where $P_n$ diverges with increasing sample size. In \eqref{eq:simulated-resp}, the mean effect $\mu_i(\cdot)$ represents the mean effect for the $i$th location, and the neighborhood term, $\mathcal N_i= \cup_{m=1}^q \mathcal N_{m,i}$, is the neighborhood of a location, $\bm s_i$ which is the union of $q$ different structured neighborhoods. Here $\mathcal N_{m,i}\equiv \mathcal N_{m(i)}$ is the $m$th-order spatial neighborhood around $\bm s_i$.
\subsection{Local Variable Selection}
In this section, we will first introduce the local variable selection method in the absence of an unobserved spatial random effect. Therefore, towards local variable selection, we first localize the covariate effect in the mean function and then select the variables at the local level. The idea of localizing the mean function is motivated by \cite{hallin2004local}. The mean function is 
\begin{equation*}
    \check{\mu}(\bm s) = \exp\left\{\sum_{k=1}^{P_n} \beta_k(\bm s) X_k(\bm s)\right\}, \ \bm s\in \mathcal{S}_n.
\end{equation*}
Using Taylor's first-order expansion, we can decompose the regression coefficient function, $\beta_k(\bm s)$, as follows:
\begin{equation}
   \beta_k(\bm s) \approx \beta_k(\bm s^{(0)}) + (\bm s - \bm s^{(0)})^\top\nabla\beta_k(\bm s^{(0)}) = \gamma_k + (\bm s - \bm s^{(0)})^\top \bm \Delta_k . 
\end{equation}
Suppose, $\bm s = (s_x, s_y)^\top$ and $\bm s^{(0)} = (s_{0,x}, s_{0,y})^\top$ then $(\bm s - \bm s^{(0)}) = [s_x - s_{0,x}, s_y - s_{0,y}]^\top$ and $\bm \Delta_k = [\nabla_x \beta_k(\bm s^{(0)}), \nabla_y \beta_k(\bm s^{(0)})]^\top = [\delta_k, \tau_k]^\top$.  The localized mean structure, $\check{\mu}_j$ for the $j$th county, $\bm s_j$ around $i$th county $\bm s^{(0)}_i$ is:
\begin{equation}\label{eq:local-mean}
\begin{split}
      \check{\mu}_j\left(\bm s^{(0)}_i|\bm \gamma, \bm \delta, \bm{\tau}\right) = \exp&\left\{\sum_{k=1}^{P_n} \gamma_k X_k(\bm s_j)+ \sum_{k=1}^{P_n} \delta_k (s_x - s_{0,x})X_k(\bm s_j)\right.\\  
      &\left. + \sum_{k=1}^{P_n} \tau_k (s_y - s_{0,y})X_k(\bm s_j) 
    \right\}.
\end{split}
\end{equation}
In (\ref{eq:local-mean}), we incorporate the localized coefficient effect arising due to predictors by incorporating three effects: $\gamma_k,$ which only quantifies the effect of the predictor at the local level, $\delta_k$ provides the effect of covariates along the horizontal direction whereas $\tau_k$ is the effect of covariates along vertical direction. We have considered the parameter vector as $\bm{\eta} = [\bm \gamma, \bm \delta, \bm{\tau}]^\top$. Thus, our localized pseudo-log likelihood w.r.t the location $\bm s^{(0)}_i$ is:
\begin{equation}\label{eq:local-loglik}
\begin{split}
    L^{(i)}(\check{\mu}(\bm \eta)) &=  \sum_{j=1}^{n_i} \left\{Y_j \log [\sigma \check{\mu}_j(\bm s^{(0)}_i | \bm \eta)]-[Y_j+1/\sigma]\log[1+\sigma \check{\mu}_j(\bm s^{(0)}_i| \bm \eta)] \right.\\
    &\left. + \log [\Gamma(Y_j + 1/\sigma)] - \log [\Gamma(1/\sigma)] - \log [\Gamma(Y_j+1)]\right\}.
\end{split}
\end{equation}
Here $n_i$ is the cardinality of the set $\mathsf{B}(h_n; \bm s^{(0)}_i) \equiv \left\{\bm s_j\in \mathcal{S}_n: \|\bm s^{(0)}_i - \bm s_j\| \leq h_n\right\}$ centred around the reference $i$th county $\bm s^{(0)}_i$. For establishing theoretical results, we assume $n_i \equiv N_n$ later. Under local mean structure in the random-effect part, the sum will be over $n_i$ counties which are in $\mathsf{B}(h_n; \bm s^{(0)}_i)$ instead of the full set of spatial sampling points on $\mathcal{S}_n$, as a result, $n$ is replaced by $n_i$ which localizes the spatial mean function. The advantage of this localization in the log-likelihood is twofold: first, it incorporates the local level spatial information in the likelihood, and second, it also incorporates the directional variation in the coefficient surface, which provides the importance of a predictor along with its direction. To select variables locally, we consider the popular SCAD penalty function, introduced by \cite{fan2001variable}. The SCAD penalty, \(\mathcal{P}_{\lambda_n}(t; \nu)\) with its tuning parameters \(\nu >2, \lambda_n >0\) will be 
\begin{equation}\label{eq:scad-penalty}
   \mathcal{P}_{\lambda_n}(t;\nu) =  \begin{cases}
        \lambda_n \lvert t \rvert; & \lvert t \rvert \leq \lambda_n,\\
        -\frac{\left(t^{2} - 2\nu\lambda_n \lvert t \rvert + \lambda_n^{2}\right)}{2(\nu-1)}; & \lambda_n < \lvert t \rvert \leq \nu\lambda_n,\\
        \frac{(\nu+1)\lambda_n^{2}}{2}; & \lvert t \rvert > \nu \lambda_n
                        \end{cases}
\end{equation}
and \(\mathcal{P}_{\lambda_n}(0;\nu) = 0\). The first derivative of \(\mathcal{P}_{\lambda_n}(t;\nu)\) is: 
\begin{equation}\label{eq:scad-derivative}
   \mathcal{P}^{\prime}_{\lambda_n}(t;\nu) =  \begin{cases}
        \lambda_n \operatorname{sgn}(t); & \lvert t \rvert \leq \lambda_n,\\
        \frac{\left(\nu\lambda_n - \lvert t\rvert\right)\operatorname{sgn}(t)}{(\nu-1)}; & \lambda_n < \lvert t \rvert \leq \nu\lambda_n,\\
        0; & \lvert t \rvert > \nu \lambda_n.
                        \end{cases}
\end{equation}
Under the SCAD penalty, the penalized local log-likelihood is
\begin{equation}\label{eq:penal-loglik}
\begin{split}
    \mathcal{L}^{(i)}(\bm \eta) =  \frac{1}{n_i}L^{(i)}(\check{\mu}(\bm{\eta}))
    - \sum_{k=1}^{P_n} \left(\mathcal{P}_{\lambda_n}(|\gamma_k|;\nu) + \mathcal{P}_{\lambda_n}(|\delta_k|;\nu) + \mathcal{P}_{\lambda_n}(|\tau_k|;\nu)\right) 
\end{split}
\end{equation}
where $\mathcal{P}_{\lambda_n}(\cdot; \nu)$ is the SCAD penalty. By maximizing the penalized log-likelihood in (\ref{eq:penal-loglik}) under the SCAD penalty in (\ref{eq:scad-penalty}), we can get the local parameter estimates
\begin{equation}\label{eq:local-parameter}
      \widehat{\bm \eta}^{(i)} = \arg \max_{\bm \eta\in \mathbb{R}^{3\cdot P_n}} \mathcal{L}^{(i)}(\bm \eta).
\end{equation}
The local estimated parameters in (\ref{eq:local-parameter}) are local SCAD penalized maximum likelihood estimates (LSPMLE), which are defined as $\widehat{\bm \eta}^{(i)}$. Based on LSPMLE, we can estimate the index set of relevant predictors. The estimated index set, 
\begin{equation}\label{eq:estimated-index}
   \widehat{\mathcal{A}}^{(i)}_n = \left\{k\in \{1,\dots, P_n\}: \text{At least one component of} \, \; \widehat{\bm \eta}_k^{(i)}\equiv (\hat{\gamma}^{(i)}_k, \hat{\delta}^{(i)}_k, \hat{\tau}^{(i)}_k)^\top \ne 0\right\} 
\end{equation}
is produced by LSPMLE and $\widetilde{\bm X}\in \mathbb{R}^{P_{0,i}}$ denotes the selected local level predictors where $P_{0,i}$ are the length of non-zero predictors w.r.t the reference $i$th county and $P_{0,i} \ll 3\cdot P_n$. For notational ease, we assume $3\cdot P_n$ as $K_n$. After standardization, the new design matrix, $\widetilde{\bm X}$, is obtained. In the next step, we employ these locally selected predictors and draw a local inference for the neighborhood.
\subsection{Local Neighborhood Inference}
Let's consider the function $\tilde{\mu}(\bm s_i^{(0)}|\cdot)$, which is the localized mean function around the $i$th county in the presence of a spatial random effect. Incorporating the locally selected variables from the set, $\widehat{\mathcal{A}}^{(i)}_n$ we can decompose the localized mean function, $\tilde{\mu}_j$ for the $j$th county, $\bm s_j$ around $i$th county $\bm s^{(0)}_i$ will become
\begin{equation}\label{eq:local-mean-random}
\begin{split}
      \tilde{\mu}_j\left(\bm s^{(0)}_i| \bm \Theta\right) = \exp&\left\{\sum_{k=1}^{P_{0,i}} \tilde{\gamma}^{(i)}_k \widetilde{X}_k(\bm s_j)+ \sum_{k=1}^{P_{0,i}} \tilde{\delta}^{(i)}_k (s_x - s_{0,x})\widetilde{X}_k(\bm s_j) \right.\\  &\left. + \sum_{k=1}^{P_{0,i}} \tilde{\tau}^{(i)}_k (s_y - s_{0,y})\widetilde{X}_k(\bm s_j) 
     + \sum_{m=1}^q \theta_m \left(\sum_{l\ne j = 1}^{n} \omega_{m,l,j} Y(\bm s_l)\right)\right\}.
\end{split}
\end{equation}
where, $\theta_m$ is the coefficient corresponding to the $m$th order spatial neighborhood around $\bm s_i$. In (\ref{eq:local-mean-random}), the parameter vector, $\bm \Theta^{(i)} = \left[\tilde{\bm \gamma}^{(i)}, \tilde{\bm \delta}^{(i)}, \tilde{\bm{\tau}}^{(i)}, \bm \theta\right]^\top$, incorporates the locally selected covariate effect and spatial random effect in the mean function. 

Before driving into the localized spatial neighborhood structure, we provide a brief overview of different types of neighborhood structure in the spatial domain, $\mathcal S_n$. We construct the $m$th type spatial weight matrix $\bm W_{m} \in \mathbb{R}^{n \times n}$ such that where the elements in $l$th row and $b$th column can be considered as, $\omega_{m,l,b}$ which is defined as:
\[
\omega_{m,l,b} = \begin{cases}
    1 &\mathrm{if}\ l\mathrm{th \ and} \ b\mathrm{th \ county\ will\ share\ its\ edge\ or\ vertices,}\\
    0 & \mathrm{Otherwise}.
\end{cases}
\]
The coefficnts $\omega_{m,l, b}$ is pre-determined spatial weight matrix where it satisfies $\omega_{m,l, b} = \omega_{m, b,l}$ for $b \ne l=1,\dots, n$ and $\omega_{m, l, l} = 0$. Based on the weights $\omega_{m,l,b}$, we can construct a spatial weight matrix $\bm{W}_m$ corresponding to the $m$th-order spatial neighborhood, where the diagonal components are zero and the matrix is symmetric.
\begin{figure}[h!]
    \centering
    \includegraphics[width=\linewidth]{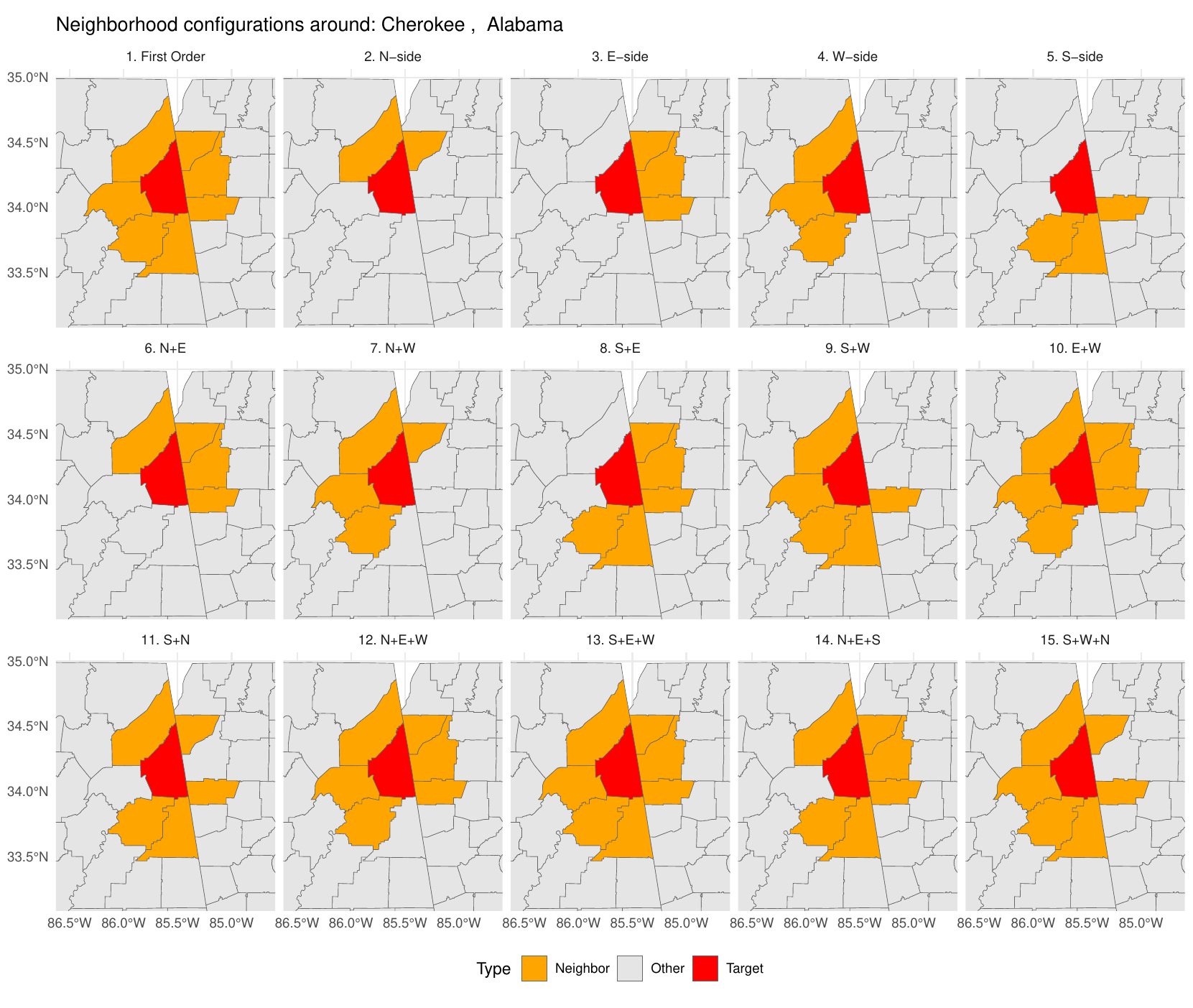}
    \caption{Examples of spatial neighborhood structures on a polygonal grid around the target county.}
    \label{fig:neighborhood-structures}
\end{figure}

Figure~\ref{fig:neighborhood-structures} illustrates a sample set of possible different spatial
neighborhood structures on areal data. From Figure~\ref{fig:neighborhood-structures}, we illustrate how we can construct different-order spatial neighborhoods around Cherokee County, Alabama. For example in the first order neigborhood method we can see in first order neighborhood structure seven counties have shared their borders with Cherokee, now if we proceed with directional anisotropy in neighborhood construction we can see in North (N) direction only two conties, in the East direction (E) there are three counties, in the West direction (W) there are three counties, in the South (S) direction there are three counties, in North-East (NE) there will be four counties and similar behavior in North-West (NW) direction. Whereas in the South-East (SE) direction, there are five counties. Eventually, around Cherokee, the neighborhood structure in the East-West (EW) direction and in the North-East-West (NEW) directions are the same, but for the boundary county, there is no guarantee that the neighborhood structure will be the same. The advantage of these different types of neighborhood structures is that we can identify which type of spatial dependence structures are present in the data in the presence of directional anisotropy. In this way, we can construct different types of neighborhood structures in the lattice data. Thus, our localized pseudo-log likelihood defined over $\mathsf{B}(h_n, \bm s_i^{(0)})$ is:
\begin{equation}\label{eq:local-loglik}
\begin{split}
    L^{(i)}\left(\tilde{\mu}(\bm \Theta^{(i)})\right) &=  \sum_{j=1}^{n_i} \left\{Y_j \log [\sigma \tilde{\mu}_j(\bm s^{(0)}_i | \bm \Theta^{(i)})]-[Y_j+1/\sigma]\log[1+\sigma \tilde{\mu}_j(\bm s^{(0)}_i| \bm \Theta^{(i)})] \right.\\
    &\left. + \log [\Gamma(Y_j + 1/\sigma)] - \log [\Gamma(1/\sigma)] - \log [\Gamma(Y_j+1)]\right\}.
\end{split}
\end{equation}
Whenever total sample size, $n$ is replaced by $n_i$ the spatial weight matrix $\bm{W}_m$ is localized and transformed into $\bm{W}_m^{(i)}$. For the post-selection inference, we consider the local mean function from (\ref{eq:local-mean-random}) for $P_{0,i}$ selected predictors included in the estimated index set \eqref{eq:estimated-index} and the local-log likelihood function from (\ref{eq:local-loglik}). We provide the inference for local parameters $\widetilde{\bm\Theta}^{(i)}=(\tilde{\bm \gamma}^{(i)}, \tilde{\bm \delta}^{(i)}, \tilde{\bm\tau}^{(i)}, \bm \theta^{(i)})^\top\in \mathbb{R}^{P_{0,i}+q}$ without using the penalty term using $p_{0,i}$ conditionally selected predictors as follows: 
\begin{equation}\label{eq:post-select-param}
\begin{split}
     \widehat{\widetilde{\bm\Theta}}^{(i)} :=& \arg\min_{\widetilde{\bm\Theta}\in \mathbb{R}^{p_{0,i} + q}} \sum_{j=1}^{n_i} \left\{Y_j \log [\sigma \tilde{\mu}_j(\bm s^{(0)}_i | \widetilde{\bm\Theta})]-[y_j+1/\sigma]\log[1+\sigma \tilde{\mu}_j(\bm s^{(0)}_i| \widetilde{\bm\Theta})] \right.\\
    &\left. + \log [\Gamma(Y_j + 1/\sigma)] - \log [\Gamma(1/\sigma)] - \log [\Gamma(Y_j+1)]\right\}.
\end{split}
\end{equation}
This estimated parameter vector $\widehat{\widetilde{\bm\Theta}}^{(i)}$ in (\ref{eq:post-select-param}) satisfies our two-fold requirements: first, local inference about locally selected predictors, and most importantly, the second one, which directional spatial dependence is inherited in the spatial pattern of firearm fatality. Thus, we can provide local inference for selected covariates and the spatial random effect.
\subsection{Computational Details}\label{ss:computational details}
In this section, we describe the computational details of LSPMLE. We capture directional anisotropy in the spatial neighborhood structure around each target county $\bm s^{(0)}_i$ in the localized spatial domain, $\mathsf{B}\left(h_n; \bm s_i^{(0)}\right)$ instead of the full $\mathcal{S}_n$. Here, we first heuristically fix the initial value of $h_n$ so that $n_i > n_0$ and $n_0$ is a predefined sample size. Now we can first describe the way of introducing five different types of neighborhood structure for a fixed county, $\bm s_j \in \mathsf{B}\left(h_n; \bm s_i^{(0)}\right)$ as follows:
\begin{equation}\label{eq:neighbor}
    \begin{aligned}
  \mathcal N_i^{\mathrm{rook}}(j) &= \{\, \bm s_\ell : \text{counties $j$ and $\ell$ share an edge}\,\},\\
  \mathcal N_i^{\mathrm{E-side}}(j) &= \{\, \bm s_\ell : \text{counties $j$ and $\ell$ share an edge along ``E'' direction}\,\},\\
  \mathcal N_i^{\mathrm{W-side}}(j) &= \{\, \bm s_\ell : \text{counties $j$ and $\ell$ share an edge along ``W'' direction}\,\},\\
  \mathcal N_i^{\mathrm{N-side}}(j) &= \{\, \bm s_\ell : \text{counties $j$ and $\ell$ share an edge along ``N'' direction}\,\},\\
  \mathcal N_i^{\mathrm{S-side}}(j) &= \{\, \bm s_\ell : \text{counties $j$ and $\ell$ share an edge along ``S'' direction}\,\}.
\end{aligned}
\end{equation}
Here, the spatial neighborhood $ \mathcal N_i^{\mathrm{rook}}(j)$ denotes the first-order localized spatial neighborhood structure around $j$th location, and in a similar way, the rest of the neighborhood structures in \eqref{eq:neighbor} denote the remaining neighborhood structure along E, W, N, and S directions. We then build additional neighborhood types by unions, e.g.,
\[
  \mathcal N_i^{\mathrm{NE}}(j)
  = \mathcal N_i^{\mathrm{N-side}}(j) \cup \mathcal N_i^{\mathrm{E-side}}(j),
  \quad
  \mathcal N_i^{\mathrm{EW}}(j)
  = \mathcal N_i^{\mathrm{E-side}}(j) \cup \mathcal N_i^{\mathrm{W-side}}(j),
\]
and more complex combinations such as
$\mathcal N_i^{\mathrm{NEW}}(j)$, $\mathcal N_i^{\mathrm{SEW}}(j)$, $\mathcal N_i^{\mathrm{NES}}(j)$, and $\mathcal N_{\mathrm{SWN}}(i)$.
Altogether, we consider $q=15$ neighborhood types:
\[
  \{\mathrm{rook},
    \mathrm{Nside},\mathrm{Eside},\mathrm{Wside},\mathrm{Sside},
    \mathrm{NE},\mathrm{NW},\mathrm{SE},\mathrm{SW},
    \mathrm{EW},\mathrm{SN},
    \mathrm{NEW},\mathrm{SEW},\mathrm{NES},\mathrm{SWN}\}.
\]
Based on all these neighborhood types, we create 15 different types of local spatial weight matrix $\bm W_m^{(i)}$.

Given a fixed $h_n = h_0$, the Step~1 SCAD estimator is indexed by the penalty level $\lambda_n>0$,
and the underlying local design matrix has dimension $n_i(h_0)\times K_n$ corresponding to the three blocks
$\bigl\{\gamma_k\bigr\}$, $\bigl\{\delta_k\bigr\}$ and $\bigl\{\tau_k\bigr\}$ for $k=1,\dots,P_n$.
For fixed $h_0$, we search for $\lambda_n$ over a data–adaptive interval, $\lambda_{\min}(h_0)
  \asymp
  \sqrt{\frac{\log 3\cdot P_n}{n_i(h_0)}},
  \
  \lambda_{\max}(h_0)
  \asymp
  \sqrt{n_i(h_0)},$
and construct a grid $\bigl\{\lambda_n^{(1)},\dots,\lambda_n^{(L)}\bigr\}
  \subset
  \bigl[\lambda_{\min}(h_0),\lambda_{\max}(h_0)\bigr]$ on the log–scale. The choice $\lambda_{\min}(h_0)$ and $\lambda_{\max}(h_0)$ is consistent with high–dimensional theory that yields sufficiently strong shrinkage so that all local coefficients can be driven close to zero for large penalties. For a given $(h_0,\lambda_n)$, we consider the localized parameters as $\widehat{\bm\eta}^{(i)}(h_0,\lambda_n)$  and based on this localized parameter using Anderson accelerated coordinate descent approach \cite{bertrand2021anderson} we can derive the localized mean function as $\widehat{\check{\mu}}_j^{(i)}(h_n,\lambda_n)$ which is a function of radius of localized spatial domain, $h_n$ and penalty parameter $\lambda_n$. 

  Because the local problems are high-dimensional and the iterative reweighted least-square (IRLS)–based coordinate descent can converge slowly, especially under strong SCAD regularization and collinear neighborhood structures, we further accelerate Step~1 using Anderson's extrapolation. The idea of local linear approximation or IRLS-based coordinate descent is actually proposed by \cite{zou2008one}, where using the weighted penalized least square method, the coefficients are iteratively updated in the outer loop, and in the inner loop, the coordinate descent takes place. Here, we expedite the inner loop using accelerated Anderson extrapolation from \cite{bertrand2021anderson}. Let's assume $\bm\eta^{(r)}$ is the set of all local coefficient parameter iterate after $r$th coordinate–descent steps for a fixed $(h_0,\lambda_n)$. After accumulating a short history $\bm\eta^{(r-\kappa+1)},\dots,\bm\eta^{(r)}$, we form the difference matrix
\[
  \bm D
  =
  \bigl[
    \bm\eta^{(r-\kappa+2)}-\bm\eta^{(r-\kappa+1)},\dots,
    \bm\eta^{(r)}-\bm\eta^{(r-1)}
  \bigr],
\]
and compute the Gram matrix $\bm G = \bm D^\top\bm D$, and solve
$\bm G\bm c = \bm 1$ to obtain normalized Anderson weights
$\bm c^\star = \bm c / (\bm 1^\top \bm c)$. The accelerated candidate is
\[
  \bm\eta^{\mathrm{acc}}
  =
  \bm\eta^{(r-\kappa+1)} + \bm D\,\bm c^\star.
\]
We then evaluate the penalized local objective $ \mathcal{L}^{(i)}(\bm\eta)$ from (\ref{eq:penal-loglik}) and accept the extrapolated iterate only if
$\mathcal{L}^{(i)}(\bm\eta^{\mathrm{acc}}) < \mathcal{L}^{(i)}(\bm\eta^{(r)})$, otherwise, retaining the unaccelerated update. This Anderson–accelerated scheme substantially reduces the number of IRLS/coordinate–descent iterations required to obtain $\widehat{\bm\eta}^{(i)}(h_0,\lambda_n)$, while preserving descent of the SCAD–penalized local likelihood. To choose $(h,\lambda_n)$, we use a \emph{weighted quasi–BIC} (WQBIC) criterion of the form
\begin{equation}\label{eq:wqBIC}
  \mathrm{WQBIC}^{(i)}(h,\lambda_n)
  =
  -\,2\,L^{(i)}\bigl(\widehat{\check{\mu}}(\bm{\eta}(h,\lambda_n))\bigr)
  + \log\bigl(n_i(h)\bigr)\, \sum_{k=1}^{K_n}
  \mathbf{1}
  \left\{
    \bigl|\widehat\eta_k^{(i)}(h,\lambda_n)\bigr| > \varepsilon
  \right\},
\end{equation}
where the second term is an estimate of the \emph{effective local model dimension}. In our implementation, the summation in the second term in \eqref{eq:wqBIC} counts how many localized covariate effects
$\bigl(\gamma_k,\delta_k,\tau_k\bigr)$ remain active at location $\bm s^{(0)}_i$ under the SCAD penalty at level $\lambda_n$. For each center $\bm s^{(0)}_i$ for a given tuning parameter $\nu$ we evaluate $\mathrm{WQBIC}^{(i)}(h,\lambda_n)$ over the grid and select
\[
  (h^\star_{n,i},\lambda^\star_{n,i})
  =
  \arg\min_{h,\lambda_n}
  \mathrm{WQBIC}^{(i)}(h,\lambda_n).
\]
The corresponding SCAD solution
$\widehat{\bm\eta}^{(i)}\bigl(h^\star_i,\lambda^\star_{n,i}\bigr)$
and the triplets $\bigl(\widehat{\widetilde{\gamma}}_k^{(i)},\widehat{\widetilde{\delta}}_k^{(i)},\widehat{\widetilde{\tau}}_k^{(i)}\bigr)$ describe the locally varying covariate effects at $\bm s^{(0)}_i$ for each predictor $k\in\widehat{\mathcal A}_n^{(i)}$, while the coefficients $\widehat\theta_m^{(i)}$ quantify the additional contribution of each directional neighborhood type $m\in\{1,\dots,q\}$ to the local log–mean of firearm fatalities. In particular, the magnitudes $|\widehat\theta_m^{(i)}|$ provide a data-driven ranking of the most influential directional neighborhood structures, while the random effect is concentrated around the county $i$ in the post-selection NB model.
\begin{algorithm}[h!]
\caption{Two-step Local NB-SCAD with Directional Neighborhood.}
\label{alg:local-nb-scad}
\begin{algorithmic}[1]
\State Based on polygon adjacency, construct different types of spatial neighborhood structure. 
\State Fit the global NB intercept-only model to get dispersion $\hat\sigma$.
\For{each county $\bm s_i$}
\State Form local covariates $\bm X_{\text{full}}=[\bm X,\ \bm X\odot(s_x-s_{x,i}),\ \bm X\odot(s_y-s_{y,i})]$.
\State \textbf{Step 1:} Minimizing WQBIC select $(h^\star, \lambda_n^\star)$ and obtain $\widehat{\mathcal{A}}_n^{(i)}$.
\State \textbf{Step 2:} Refit unpenalized local NB on $\widetilde{\bm X}$; get neighborhood coefficients $\hat\theta$.
\State Select the neighborhood structure based on the largest value of $\hat{\theta}_m$.
\EndFor
\end{algorithmic}
\end{algorithm}
In this Algorithm~\ref{alg:local-nb-scad}, we have summarized the computational details of local variable selection and spatial neighborhood selection.
\section{Theoretical Results}\label{s:theory}
In this section, we will establish the existence of a local maximizer \eqref{eq:penal-loglik} and the infill asymptotics of variable selection consistency. Here, the reason behind the infill asymptotics is that at the time of local variable selection, we only consider the spatial region, $\mathsf{B}_i$, which is the subset $\mathcal{S}_n$ of a fixed radius $h_0$. Now, while we study the large sample properties of variable selection consistency, the sampling points, $n_i\equiv N_n$, within this fixed region will increase, and thus this subregion becomes dense, and we will study the large sample properties. For the mathematical details of infill asymptotics in spatial data please go through \cite{lahiri1996inconsistency}. Now let's consider out of $K_n = 3\cdot P_n$ variables only $P_0$ variables are relevant and we arrange them consecutively such that first $P_0$ predictors are relevant which are stored into the set $\mathcal{A}_n^{(i)}$ and the remaining irrelevant predictors are stored into the set $\mathcal{B}_n^{(i)}$. Thus $[[\mathcal{A}_n^{(i)}]] = P_0$ and $[[\mathcal{B}_n^{(i)}]] = P_1$ such that $P_0+P_1 = 3\cdot P_n = K_n$ where $[[\cdot]]$ represents the cardinality of a set. Let's consider $\bm \eta^0_i \equiv \bm \eta_0$ be the true regression coefficient, where the first $P_0$ coefficients are nonzero and the last $P_1$ components are zero. For the notational simplicity we assume $\check \mu_j(\bm s_i^{(0)} | \bm \eta)$ as $\check \mu_{ji}(\bm \eta) = \exp\{\bm z_j^\top \bm \eta\}$ where $\bm z_j = (\bm X_j, (s_{x,j} - s_{x,i}) \bm X_j, (s_{y,j} - s_{y,i}) \bm X_j)^\top\in \mathbb{R}^{K_n}$ for all $j \in \{1,\dots, N_n\}$. Then the score function and the Fisher information matrix are respectively:
\begin{equation}\label{eq:score-fisher}
    \begin{aligned}
        \bm\omega_n^{(i)}(\bm\eta) &= \sum_{j=1}^{N_n} \left(Y_j -\frac{(\sigma Y_j + 1)\check{\mu}_{ji}(\bm \eta)}{(1+\sigma \check{\mu}_{ji})}\right)\bm z_j= \left[\bm\omega_{0n}^{(i)}(\bm\eta), \bm\omega_{1n}^{(i)}(\bm\eta)\right]^\top\\
        \bm \Omega_n^{(i)}&=\sum_{j=1}^{N_n} \left(\frac{\check{\mu}_{ji}(\bm \eta)}{(1+\sigma \check{\mu}_{ji})}\right)\bm z_j \bm z_j^\top=\begin{bmatrix}
            \bm\Omega_{00n}^{(i)}(\bm\eta) & \bm\Omega_{01n}^{(i)}(\bm\eta)\\
            \bm\Omega_{10n}^{(i)}(\bm\eta) & \bm\Omega_{11n}^{(i)}(\bm\eta)
        \end{bmatrix}
    \end{aligned}
\end{equation}
Thus like \eqref{eq:score-fisher} we can separate the score vector and fisher information matrix for relevant and irrelevant variables where $\bm\omega_{0n}^{(i)}(\bm\eta) \in \mathbb{R}^{P_0}, \bm\omega_{1n}^{(i)}(\bm\eta) \in \mathbb{R}^{P_1} $. Similarly, the partitioned components of the Fisher Information matrix are respectively, $\bm\Omega_{00n}^{(i)}(\bm\eta) \in \mathbb{R}^{P_0\times P_0}$, $\bm\Omega_{01n}^{(i)}(\bm\eta) \in \mathbb{R}^{P_0\times P_1}$, $\bm\Omega_{10n}^{(i)}(\bm\eta) \in \mathbb{R}^{P_1\times P_0}$, and $\bm\Omega_{11n}^{(i)}(\bm\eta) \in \mathbb{R}^{P_1\times P_1}$. Now we establish the important theoretical results regarding the existence of a local-maximizer of \eqref{eq:penal-loglik} and establish its convergence rate. After that, under some regularity conditions, we establish the consistent sparsity selection property of LSPMLE.
\begin{assumption}\label{ass:differentiablity-criterion}
    The criterion $\mathcal L_n^{(i)}(\bm\eta)$ is twice continuously differentiable in a neighborhood of $\bm\eta^0$ and for some fixed positive constant, $c_0 > 0$ we have, $\min_{1\le k\le P_0}|\eta_k^0|\ge c_0>0$.
\end{assumption}
\begin{assumption}\label{ass:knassum}
    The number of predictors will satisfy $\alpha_n = \sqrt{\frac{K_n}{N_n}} \to 0$ whenever $N_n \to \infty$. 
\end{assumption}
\begin{assumption}\label{ass:penalty-parameter}
    The penalty parameter $\lambda_n$ would satisfy
    \[
\alpha_n=o(\lambda_n),\qquad
\sqrt{K_n}\alpha_n=o(\lambda_n),\qquad
K_n\alpha_n^2=o(\lambda_n).
\]
whenever, $N_n \to \infty$.
\end{assumption}
\begin{assumption}\label{ass:fisher}
    The Fisher information matrix in \eqref{eq:score-fisher} will satisfy
    \[
    \lim\inf_{N_n \to \infty}\inf_{\bm \phi \in \mathbb{R}^{P_{0}}, \|\bm \phi\| = 1}\frac{1}{N_n}\bm \phi^\top \bm \Omega_{00n}^{(i)} \bm \phi = \Lambda_n> 0.
    \]
    for all non-zero $P_0$ dimensional non-zero unit-vector $\bm \phi$.
\end{assumption}
\begin{assumption}\label{ass:score}
The score vector in \eqref{eq:score-fisher} will satisfy
    \[
\left\|\frac{1}{N_n}\bm\omega_{0n}^{(i)}(\bm\eta^0)\right\| = O_P(\alpha_n).
\]
\end{assumption}
\begin{assumption}\label{ass:thirdderiv}
   There exists $r>0$ and a constant $C_3>0$ such that, uniformly for all
$k,l,m\in\{1,\dots,K_n\}$ and for all $\tilde{\bm\eta}$ with
$\|\tilde{\bm\eta}-\bm\eta^0\|\le r$ we have,
\[
\left|
\frac{1}{N_n}
\frac{\partial^3 L^{(i)}(\tilde{\bm\eta})}
{\partial\eta_k\partial\eta_l\partial\eta_m}
\right|
\le
C_3.
\]
\end{assumption}
\begin{assumption}\label{ass:bounded-z}
\begin{enumerate}
    \item There exists a fixed, $C_z>0$ such that \(\max_{1\le j\le N_n}\|\bm z_j\|\le C_z<\infty,\) so in particular $|z_{jk}|\le \|\bm z_j\|\le C_z$ for all $j,k$.
    \item There exists a fixed positive constant $\tilde{C}$ such that, 
\(\max_{1\le j \le N_n}\left|\frac{\exp(\bm z_j^{\top}\bm\eta_0)}
     {(1+\exp(\bm z_j^{\top}\bm\eta_0))^2}\right|<\tilde{C}.
\)  
\item There exists a fixed positive constant $C_\mu> 0 $ such that \(\max_{1\le j\le N_n}\left|\frac{\tilde{\mu}_j^0}{1+\sigma \tilde{\mu}_j^0}\right| < C_\mu\).
\end{enumerate}

\end{assumption}
\begin{assumption}\label{ass:penalty}
For each $k\le P_{0}$ with $\eta_k^0\neq 0$, the penalty $\mathcal P_{\lambda_n}(\cdot)$
is twice differentiable at $|\eta_k^0|$ and assume,
\[
a_n:=\max_{1\le k\le P_0}\mathcal P'_{\lambda_n}(|\eta_k^0|) = o(\alpha_n),\qquad
b_n:=\max_{1\le k\le P_0}\bigl|\mathcal P''_{\lambda_n}(|\eta_k^0|)\bigr| = o(\alpha_n).
\]
\end{assumption}
Now we give some insights into the assumptions. A-\ref{ass:differentiablity-criterion} assumes the differentiability of the loss function and the minimum of a non-zero coefficient bounded away from a given positive constant $c_0$. It's a conventional assumption like the ``beta-min'' condition in LASSO problems. A-\ref{ass:knassum} considers that the number of predictors, $K_n$ may be divergent but in slower rate than the rate of increase of sample size, $N_n$, under infill asymptotics. This assumption actually connects the two rates: one is the rate of increase of sampling points in a bounded spatial domain, and the other is the increase in the number of predictors. A-\ref{ass:penalty-parameter} is a conventional assumption regarding the penalty parameter, $\lambda_n$ which is similar to the \cite{fan2001variable}. A-\ref{ass:fisher} is a similar assumption to Assumption~B and A-\ref{ass:thirdderiv} is like Assumption~C in \cite{fan2001variable}. A-\ref{ass:score} implies that the score function vanishes in the true regression coefficients, but the bound gives insight into the convergence rate. Likewise, the traditional regularity conditions A-\ref{ass:bounded-z} provide the restriction on the design matrix. In A-\ref{ass:penalty} the term $b_n \to 0$, but we have specified the rate which is mentioned in Theorem~1 of \cite{fan2001variable} but in our case the restriction $a_n$ differs from \cite{fan2001variable}'s approach they assume that $a_n = O(N_n^{-1/2})$ where as in our case $a_n$ is converging to zero in slower rate than that of \cite{fan2001variable}. Finally, \eqref{ass:sparsity-penalty} is a similar assumption mentioned in Equation~3.5 in \cite{fan2001variable}. 
\begin{theorem}\label{thm:sparsity}
Under the regularity conditions in A-\ref{ass:differentiablity-criterion}- A-\ref{ass:penalty}
\begin{itemize}
    \item[(i).]  There exists a local maximizer $\widehat{\bm\eta}^{(i)}$ such that $\|\widehat{\bm \eta^{(i)}} - \bm{\eta}_i^0\| = O_P(\alpha_n)$.
    \item[(ii).] Additionally for some $\kappa_0>0$ if 
    \begin{equation}\label{ass:sparsity-penalty}
        \lim\inf_{N_n \to \infty}\lim\inf_{t \to 0^+} \frac{\mathcal P'_{\lambda_n}(t)}{\lambda_n}\ge \kappa_0 > 0
    \end{equation}
Then for a given reference location $\bm s_i^{(0)}$ we have,
\[
\mathbb P\Bigl(\widehat\eta^{(i)}_k=0\ \text{for all }k\in\mathcal{B}_n^{(i)}\Bigr)\to 1.
\]
\end{itemize}
\end{theorem}
Thus, our theoretical results in Theorem~\ref{thm:sparsity} have established that there exists a local maximizer in \eqref{eq:penal-loglik} with $\alpha_n$-convergence rate, which is slower than $\sqrt{N_n}$ convergence rate, and secondly, we establish that a $\alpha_n$ convergent local maximizer depicts consistent behavior in sparsity selection. The take-home message from this theoretical result is that under infill asymptotics, we can establish the existence of $\alpha_n$ convergent and consistent sparsity selector, LSPMLE.

\section{Simulation Study}\label{s:simulation}
\begin{figure}[h!]
    \centering
    \includegraphics[width=0.6\linewidth]{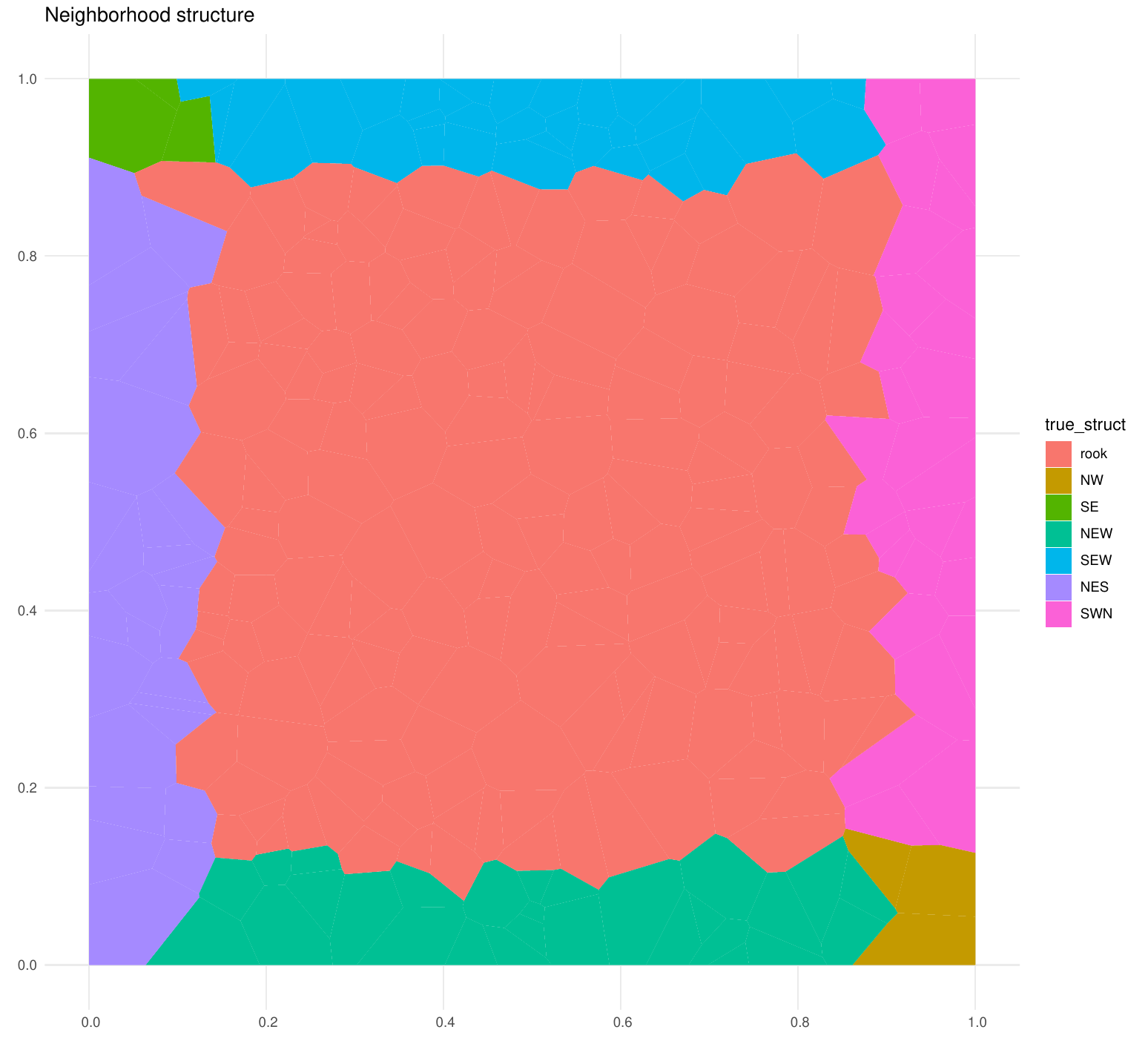}
    \caption{Hidden Spatial Neighborhood Structure used for simulated data.}
    \label{fig:spatial-neighborhood-sim}
\end{figure}

\begin{figure}[h!]
    \centering
    \includegraphics[width=0.7\linewidth]{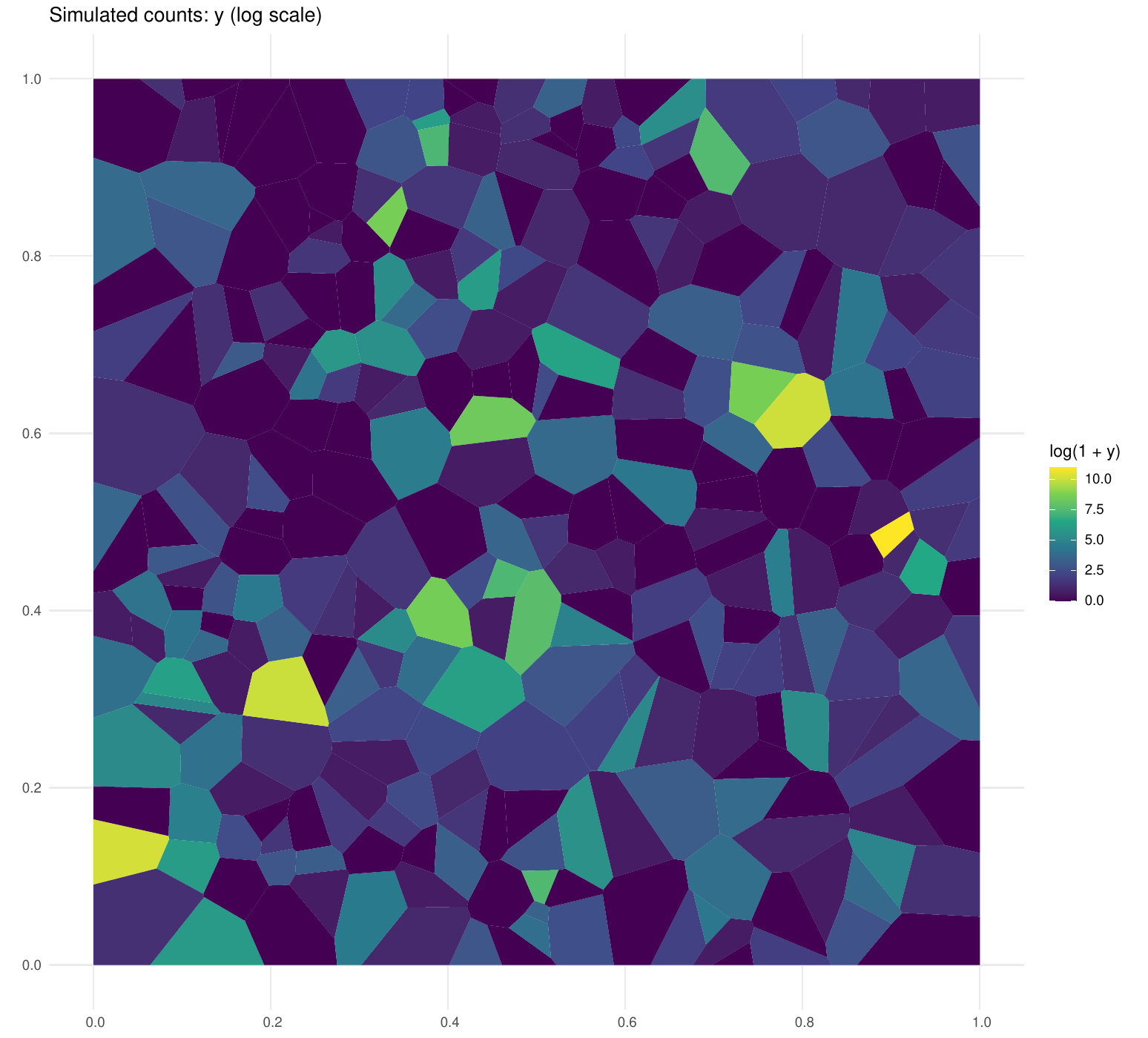}
    \caption{Spatial Distribution of Response in Simulation Study.}
    \label{fig:spatial-response-sim}
\end{figure}

We simulate spatial count data on an irregular lattice generated by Voronoi tessellation over the unit square $[0, 1]^2$. A total of $n$ polygons are constructed from uniformly sampled spatial locations, resulting in heterogeneous polygon sizes and shapes that resemble realistic areal units. Spatial adjacency is defined using \eqref{eq:neighbor} as fifteen predefined neighborhood structures, including horizontal, vertical, diagonal, and composite directions. In Figure~\ref{fig:spatial-neighborhood-sim}, we have illustrated the latent spatial neighborhood structure. We have divided the entire spatial surface into seven sub-regions, where each sub-region in Figure~\ref{fig:spatial-neighborhood-sim} reveals different types of spatial neighborhood structure. Each spatial unit is assigned a single true neighborhood structure based on its location (interior, edge, or corner), ensuring boundary-aware spatial dependence. The covariate space consists of $P_n=800$ predictors, where only the first 10 covariates are truly active. Their regression coefficients vary spatially according to smooth and piecewise-defined coefficient surfaces. The coefficient surfaces are defined as follows:
\[
\beta_1(\bm s)=
\begin{cases}
\sqrt{2}, & (s_x-0.5)^2+(s_y-0.5)^2 \le 0.2,\\
1,        & 0.2 < (s_x-0.5)^2+(s_y-0.5)^2 \le 0.3.
\end{cases}
\]
\[
\beta_2(\bm s)=
\begin{cases}
\sqrt{2}, & |s_x-0.5|+|s_y-0.5| \le 0.2,\\
-1,       & 0.2 < |s_x-0.5|+|s_y-0.5| \le 0.3.
\end{cases}
\]
\[
\beta_3(\bm s)=
\begin{cases}
\sqrt{2}, & s_x \le 0.1,\\
1,        & 0.1 < s_x \le 0.9,\\
0,        & \text{o.w.}
\end{cases}
\beta_4(\bm s)=
\begin{cases}
\sqrt{2}, & s_y \le 0.1,\\
1,        & 0.1 < s_y \le 0.9,\\
0,        & \text{o.w.}
\end{cases}
\]
\[
\beta_5(\bm s)=
\begin{cases}
\sqrt{2}, & 0.2 \le s_x \le 0.9,\ \ 0.2 \le s_y \le 0.9,\\
0,        & \text{o.w.}
\end{cases}
\]
\[
\begin{aligned}
    \beta_6(\bm s) \;=\;
\begin{cases}
\;\;1, & 0\le s_x,s_y<\tfrac12,\\[2pt]
-1, & s_x\ge \tfrac12,\; s_y\ge \tfrac12,\\[2pt]
\;\;0, & \text{otherwise}. \end{cases}\; \beta_7(\bm s) \;=\;
\begin{cases}
\;\;1, & s_x<\tfrac12,\; s_y<\tfrac12,\\[2pt]
\;\tfrac12, & \tfrac12\le s_x<\tfrac34,\; \tfrac12\le s_y<\tfrac34,\\[2pt]
\;\;0, & s_x\ge \tfrac34,\; s_y\ge \tfrac34,\\[2pt]
-1, & \text{otherwise}.
\end{cases}\\
\beta_8(\bm s) \;=\; \mathbb{I}\{\,s_x\le\tfrac12\,\}, \beta_9(\bm s) \;=\; \mathbb{I}\{\,s_y\le\tfrac12\,\}, \beta_{10}(\bm s) \;=\;
\begin{cases}
\sqrt{2}, & s_x\le\tfrac12,\; s_y\le\tfrac12,\\[2pt]
1, & s_x\le\tfrac12,\; s_y>\tfrac12,\\[2pt]
0, & s_x>\tfrac12,\; s_y>\tfrac12,\\[2pt]
0, & s_x>\tfrac12,\; s_y\le\tfrac12.
\end{cases}
\end{aligned}
\]
To enforce structured sparsity, the design matrix is constructed so that each covariate is nonzero only at locations where its true coefficient is nonzero, yielding a spatially aligned high-dimensional design. Spatial dependence is introduced through a structured spatial autoregressive random effect. Suppose $\mathcal{N}_{m(i)}$ is the $m$th directional latent spatial neighborhood assumed for the $i$th reference polygon, \(\bm s_i^{(0)}\). In a simulation study for each polygon, we fix a specific type of neighborhood structure. So, for example, if we set $i=1$ and set $m(i) = 2$, which implies the spatial dependence for the polygon with id $1$ exists in the north and west directions, as shown in Figure~\ref{fig:spatial-neighborhood-sim}. As a result, the first row of the spatial weight matrix $\widetilde{\bm W}$ only consists of the non-zero numbers where the neighboring polygons share their border with the first polygon. Suppose the spatial random effect, $g_j$ corresponding to the $j$th polygon in simulated data included in the local spatial domain, $\mathsf{B}\left(h_0, \bm s_i^{(0)}\right)$, and for a given $m \in \{1,\dots, q\}$ is defined as
\[
g_j = \rho \sum_{l=1}^n \omega_{m,j,l} \ g_l + \epsilon_j,
\]
where the nugget effect $\epsilon_k \overset{iid}{\sim} N(0, \tilde{\sigma}^2)$. We set the intercept term $\beta_0 = 1, \tilde{\sigma} = 0.6, \rho = 0.4$ and the overdispersion parameter, $\sigma = 0.5$. Spatial dependence is restricted to neighbors belonging to each location's assigned neighborhood structure. The response is simulated in the following way: 
\[
\begin{aligned}
    &\left(I_n - \rho \widetilde{\bm W}\right) \bm g = \bm\epsilon, \ \epsilon_j \overset{iid}{\sim} N(0, \tilde{\sigma}^2),\\
    &\mu_j = \exp\left\{\beta_0 + \sum_{k=1}^{P_n} \beta_k(\bm s_j) X_{k,i}+g_j\right\},\
    Y_j \vert \bm X_j \sim \mathrm{NB}(\mu_j, \sigma).
\end{aligned}
\]
Specifically, the mean is defined through a log-linear model that incorporates an intercept, spatially varying covariate effects, and a structured spatial random effect. The dispersion parameter is fixed, and the final counts are sampled accordingly. 
\begin{table}[h!]
\centering
\caption{Simulation results for $P_n = 800$ under the spatial negative binomial model.}
\label{tab:simulation-perform}
\small
\begin{tabular}{ccccccc}
\hline
$n$ & $P_n$  & Method & $\mathrm{TGVSP}$ & TLVSP \\
\hline
200 & 800 & Local-SCAD  & 0.4 & 0.5000000 \\
 &  &  Global SCAD       & 0.9 & -- \\
& &  Global AL       & 0.9 & -- \\
\hline
 300& 800 & Local-SCAD  &  0.8 & 0.7013889 \\
 & & Global SCAD         & 0.9 & -- \\
 & &  Global AL       & 1.0 & -- \\
\hline
400 & 800 & Local-SCAD & 0.8 & 0.7142857 \\
 & &  Global SCAD         &  0.9 & -- \\
& & Global AL       & 1.0 & -- \\
\hline
500 & 800 &  Local-SCAD &  1.0 & 0.6279762 \\
&  & Global SCAD         &  1.0 & -- \\
&  & Global AL       &  1.0 & -- \\
\hline
800 & 800 & Local-SCAD &  1.0 & 0.8678571 \\
& & Global SCAD         &  0.8 & -- \\
& & Global AL       &  1.0 & -- \\
\hline
\end{tabular}
\end{table}
Figure~\ref{fig:spatial-response-sim} reveals the spatial dependence in the response. This design allows simultaneous evaluation of variable selection, spatial adaptivity, and neighborhood-structure recovery under realistic irregular spatial settings. Suppose we have $M$ reference polygons, $\left\{\bm s^{(0)}_1,\dots, \bm s^{(0)}_M\right\}$. In this comparison, we have focused on two metrics: one is the true global variable selection probability (TGVSP), and the other is the true local variable selection probability (TLVSP). For the TGVSP, we consider two sets:
\[
\mathcal{E}_n \equiv \left\{p \in \{1,\dots, K_n\}: \text{there exists at least one }\ \bm s_i \ \mathrm{such\ that}\ \beta_p(\bm s_i) \ne 0\right\}
\]
and for global methods such as for Global SCAD, and AL method, the set, 
\[
\widehat{\mathcal{J}}_n \equiv \left\{p \in \{1,\dots, P_n\}:  \widehat{\beta}_p \ne 0\right\}
\]
 selects the non-zero variables, and for local scad we consider the set, $\widehat{\mathcal{A}}_n^{(i)}$ in \eqref{eq:estimated-index}. Then, the metric TGVSP for the global methods is defined as follows, 
\[
\mathrm{TGVSP} = \frac{[[\mathcal{J}_n \cap \widehat{\mathcal{J}}_n]]}{P_{0}},
\]
and for the local methods, the TGVSP becomes 
\[
\mathrm{TGVSP} = \frac{[[\mathcal{J}_n \cap (\bigcup_{i=1}^M\widehat{\mathcal{A}}_n^{(i)}]]}{P_{0}},
\]
 where $[[\cdot]]$ quantifies cardinality of a set, and $P_{0}$ indicates the number of non-zero variables. Next, we define the true TLVSP. Let $\mathcal{A}_n(\bm s_i^{(0)}) \equiv \{p\in \{1,\ldots, P_n\}: \beta_p\left(\bm s^{(0)}_i\right) \ne 0\}$ then 
\[
\mathrm{TLVSP} = \frac{1}{M}\sum_{i=1}^M \frac{[[\mathcal{A}_n(\bm s^{(0)}_i)\cap \widehat{\mathcal{A}}_n^{(i)}]]}{[[\mathcal{J}_n(\bm s^{(0)}_i)]]}.
\]
In Table~\ref{tab:simulation-perform} we compare the performance of LSPMLE with other penalized maximum likelihood estimates. From Table~\ref{tab:simulation-perform}, we observe the supremacy of LSPMLE in TGVSP and TLVSP over the global SCAD and AL with increasing sample size. Specifically, its local variable selection ability outperforms the other methods. 
\section{Firearm Fatality Data Analysis}\label{s:real-data}
\begin{figure}[h!]
    \centering
    \includegraphics[width=\linewidth]{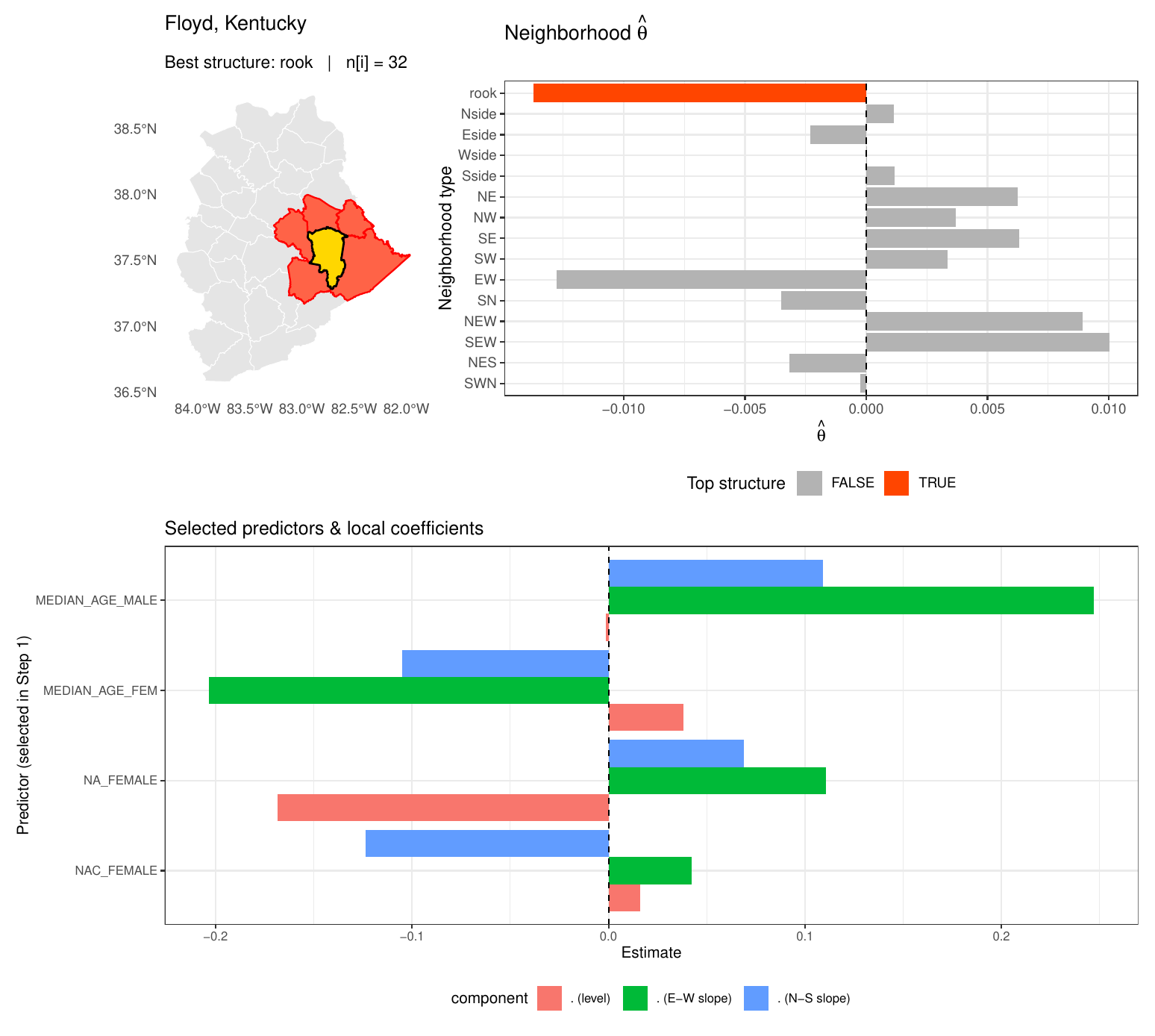}
    \caption{Local neighborhood structure and selected covariates for Floyd County.}
    \label{fig:floyd}
\end{figure}
\begin{figure}[h!]
    \centering
    \includegraphics[width=\linewidth]{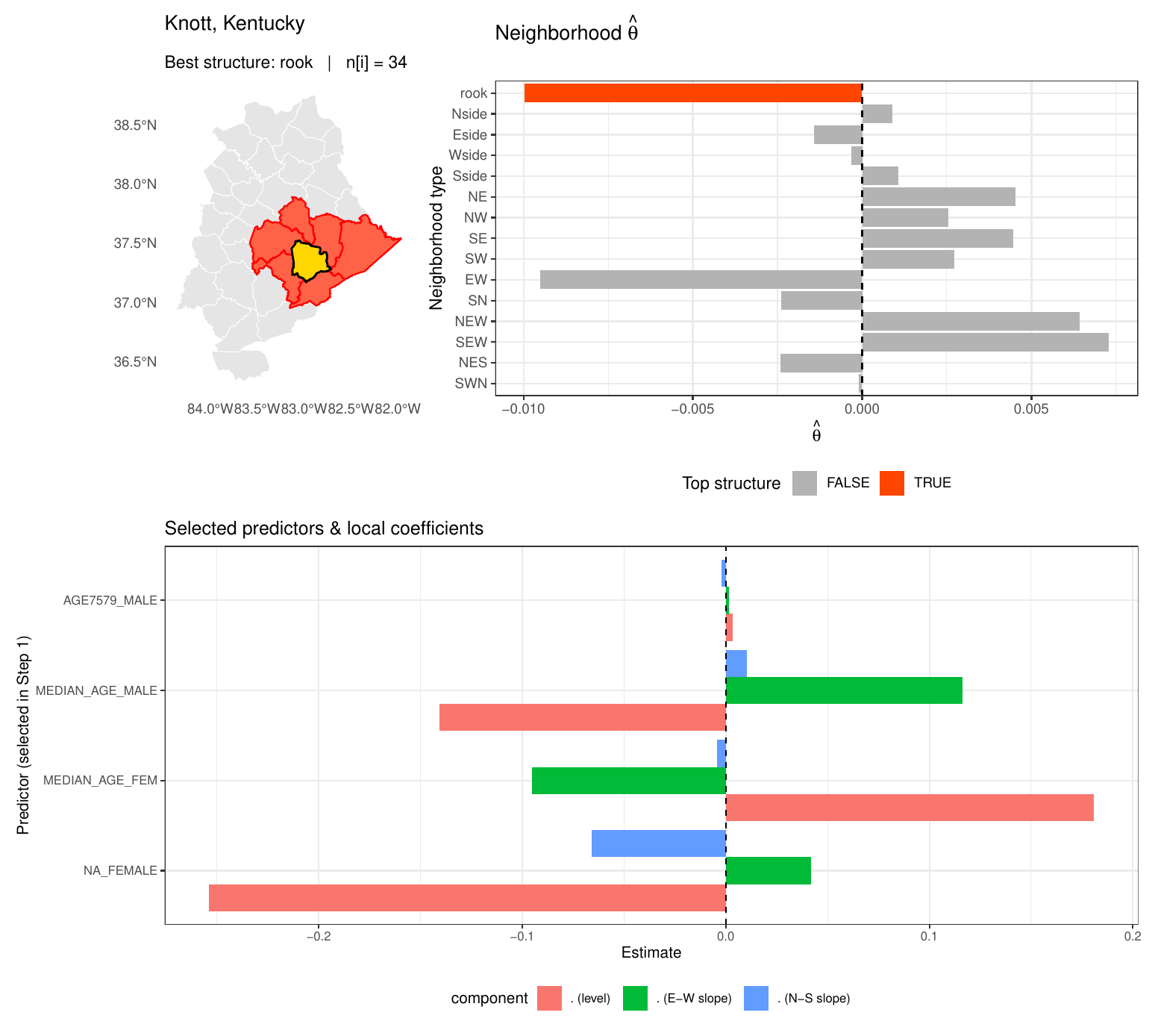}
    \caption{Local neighborhood structure and selected covariates for Knott County.}
    \label{fig:knott}
\end{figure}
\begin{figure}[h!]
    \centering
    \includegraphics[width=\linewidth]{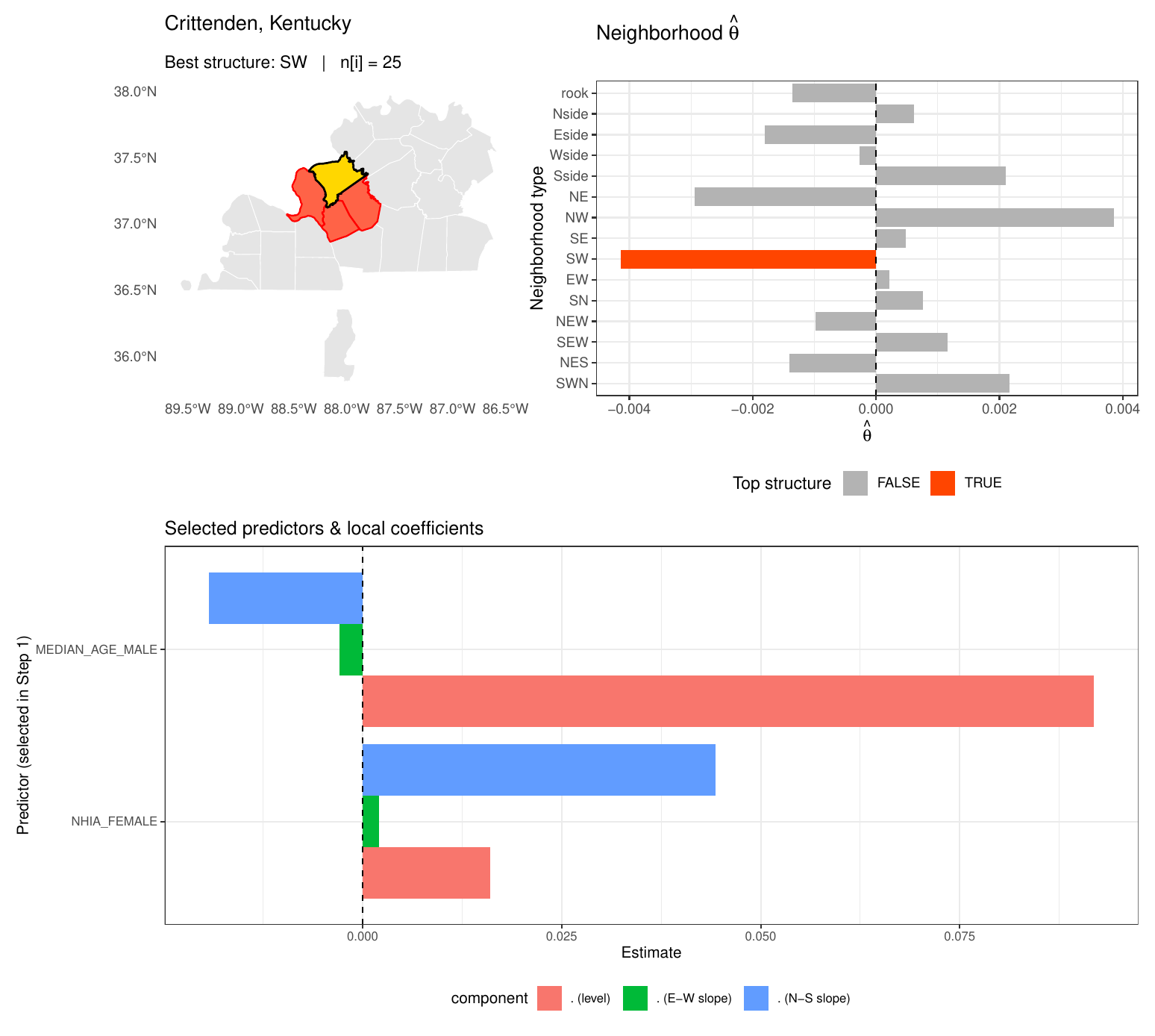}
    \caption{Local neighborhood structure and selected covariates for Crittenden County.}
    \label{fig:crittenden}
\end{figure}

\begin{figure}[h!]
    \centering
\includegraphics[width=0.8\linewidth]{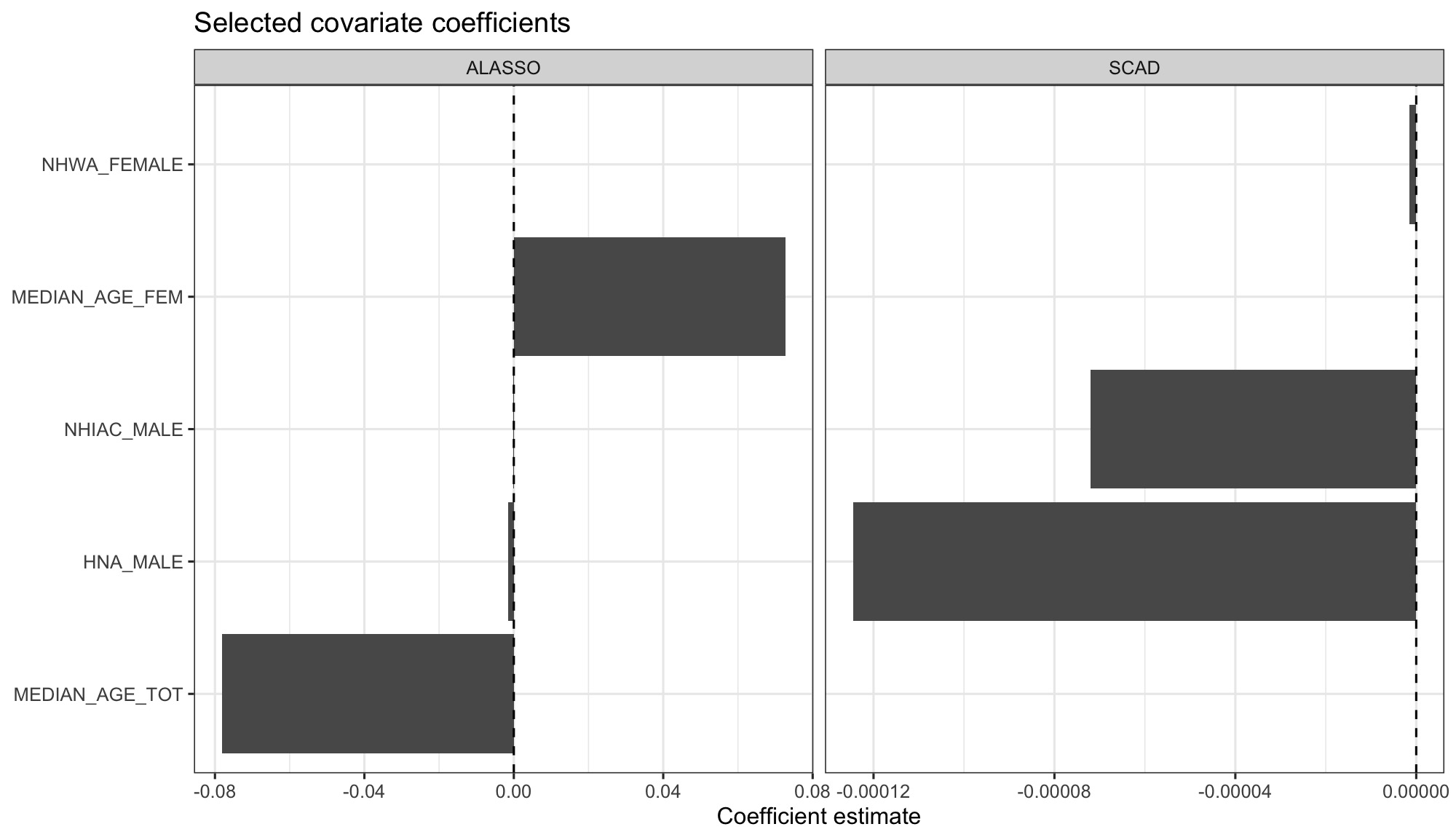}
    \caption{Selected covariates by global Methods.}
    \label{fig:global-vs}
\end{figure}
\begin{figure}[h!]
    \centering
    \includegraphics[width=\linewidth]{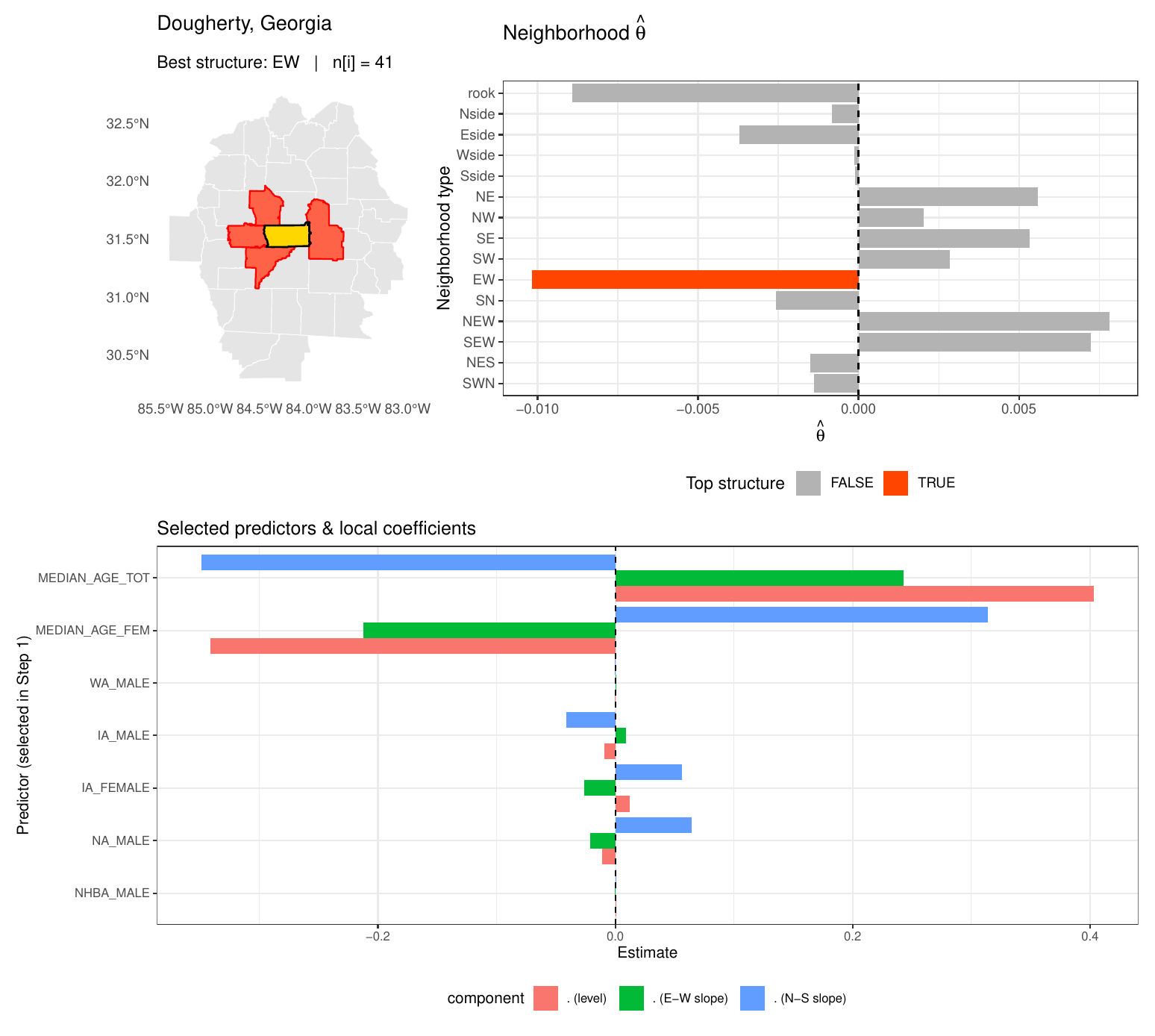}
    \caption{Local neighborhood structure and selected covariates for Dougherty County.}
    \label{fig:dougherty}
\end{figure}
\begin{figure}[h!]
    \centering
    \includegraphics[width=\linewidth]{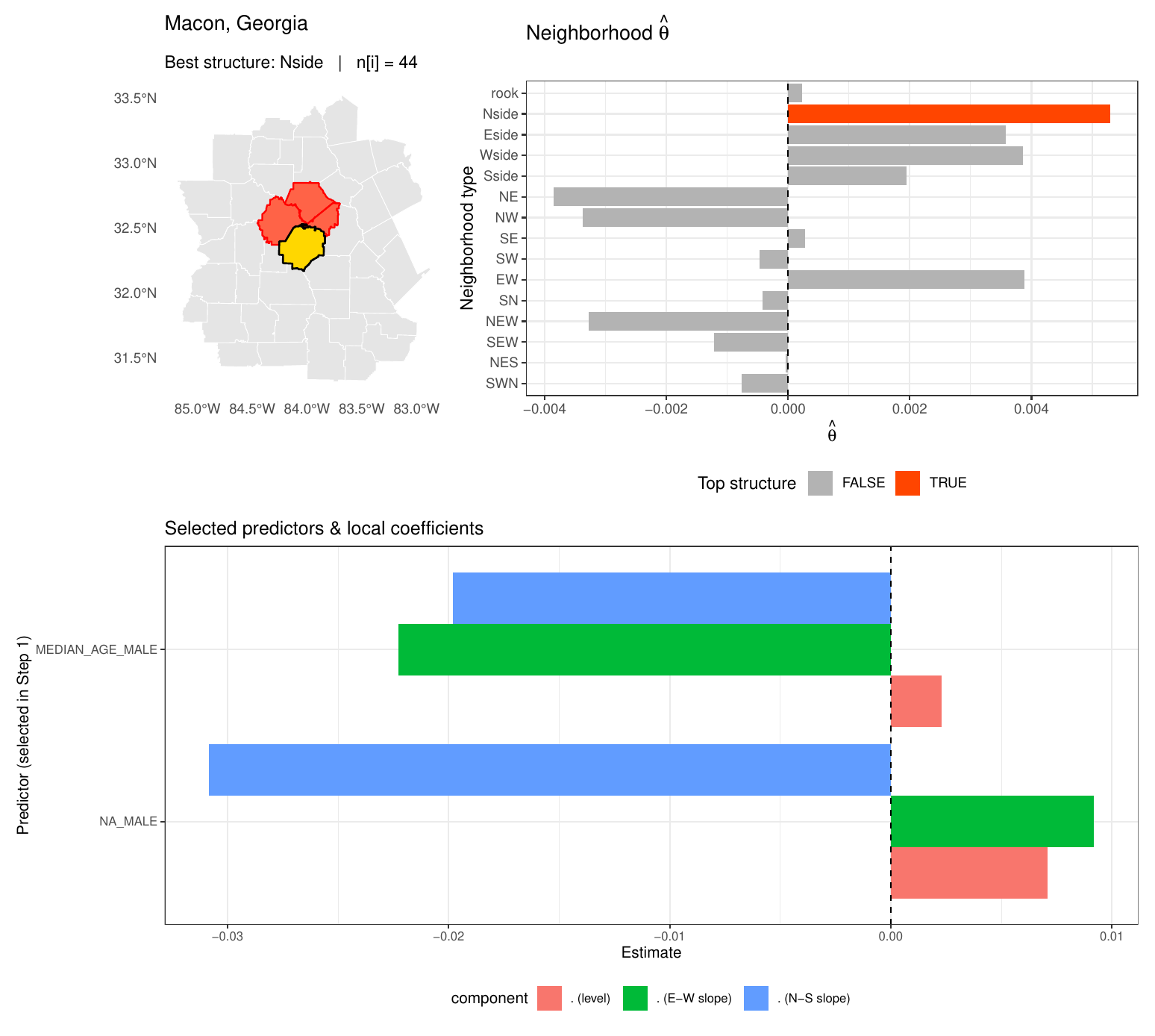}
    \caption{Local neighborhood structure and selected covariates for Macon County.}
    \label{fig:macon}
\end{figure}

In this section, we summarize the local variable selection and select the suitable neighborhood structure as a post-selection inferential variant. In 2024, Floyd County, Kentucky, had several major gun-related events that showed how gun violence may happen in diverse situations. In March 2024, a Kentucky State Police officer shot and killed 35-year-old Glenn Edward Bays from Harold during a traffic stop mentioned in the news report in \url{https://mountain-topmedia.com/ksp-reveals-more-details-about-fatal-floyd-traffic-stop-shooting/}. Authorities said he reached for a gun in his car. In Pikeville, a different incident happened later in October 2024 at a parking garage. It involved a local adolescent named Brandon Poston, mentioned in the news report \url{https://www.wymt.com/2025/11/20/teen-sentenced-parking-garage-shooting-case/}. The victim of the shooting lived after getting treatment for a gunshot wound, but the event had serious legal ramifications. Poston was later sentenced to eight years in jail for second-degree assault and having a gun as a minor. These events show both gun violence connected to law enforcement and gun violence between people in the county in the same year. According to the news report, \url{https://www.kentuckystatepolice.ky.gov/news/p13-9-19-2024} in 2024, two different events in Knott County, Kentucky, led to several deaths due to firearms. In April, there was a tragic shooting in Dwarf, which led to a murder charge against a suspect. This shows how bad gun violence is in the neighborhood. A police-involved shooting happened later in September after a long confrontation with Travis Pratt, who was killed in the incident. While public reports don't go into much information beyond these incidents, both cases added to the county's gun death toll for the year and show that there is both criminal and police enforcement-related gun violence. These are two distinct things that happened in Floyd and Knott counties. From Figure~\ref{fig:floyd} and Figure~\ref{fig:knott}, we can comment that the first-order spatial random effect structure works effectively. From the point of view of local variable selection the median ages of male, female, Native Hawaiian, and Pacific Islander residents alone and female populations (NA\_Female) are commonly selected for both counties, Floyd and Knott, whereas Native Hawaiian and Pacific Islander residents alone and in combination with female populations (NAC\_Female) are selected for Floyd. In contrast with Floyd, the Male population between $75-79$ is selected for Knott. In Figure~\ref{fig:crittenden}, we observe that the hidden spatial random effect prefers the spatial dependency in the South and West direction. Besides the Median Age of the male population non-Hispanic American Indian and Alaska Native female population (NHIA\_Female) is also a potential reason behind this firearm fatality. In Figure~\ref{fig:dougherty} we present the selected predictors and selected neighborhood structure for Dougherty County. We observe that for Firearm Fatality incidence in Dougherty median age of the total population, the median age of females, the white male population (WA\_Male), the American Indian Male and Female population (IA\_Male, IA\_Female), the Native Hawaiian male population (NA\_Male), Non-Hispanic black African American male population are selected (NHBA\_Male). We also detect that the spatial dependence in this local neighborhood is distributed in the East-West direction. From Figure~\ref{fig:macon}, we detect that the spatial dependence in the neighborhood of Macon County is distributed along the North direction, and median male age and Native Hawaiian male (NA\_Male) population are selected. In Figure~\ref{fig:global-vs} we visualize the list of selected variables by three global methods, where the first column indicates the selected variables by AL, and the second column provides the list of selected variables by SCAD penalty. We can visualize that AL selects the median age of the male and female population, the median age of the total population, and, in spatial dependence, AL has suggested that the dependence is distributed mostly in the North direction. The second column provides the selected variables by SCAD, where the non-Hispanic white female population (NHWA \_Female), the non-Hispanic American Indian male population (NIHAC\_Male), and the Alaska Native female population are selected (NAC\_Female). Thus, we can select the variables locally for six counties and also provide the directional heterogeneity in spatial neighborhood structure.

\section{Discussion}\label{s:conclusion}
Firearm mortality in the United States demonstrates significant regional heterogeneity and substantial demographic disparities, highlighting the necessity for statistical approaches that may pinpoint locally pertinent risk factors instead of depending on globally averaged effects. In this study, we devised a localized variable selection and inference framework specifically designed for spatial regression contexts, inspired by the examination of firearm fatality data. The proposed method combines spatially weighted local regressions with SCAD penalization. This approach lets you choose predictors locally, considering the directional variation in the latent neighborhood structure of gun violence data in the southeast US.

The two-step process of local variable selection, refitting, and inference solves a major problem with most of the current research on spatial variable selection, which usually just looks at global selection or does not provide effective insights about local post-selection inference for the latent spatial neighborhood structure. Our method lets us find both locally important covariates and good neighborhood structures at the same time, which helps us better understand how spatial risk works. A simulation study reveals that the suggested strategy works well for selecting true non-zero variables. Our LSPMLE not only provides insight into county-specific variable selection but also determines the suitable latent random effect structure. Our theoretical results justify the theoretical relevance of LSPMLE in sparsity selection. There are still many avenues for future study to explore. It makes sense to move on to the following phases, which include adding extensions to fully generalized linear models for count outcomes, adding spatial random effects to a unified penalized likelihood framework, and doing theoretical research with higher spatial resolution. Furthermore, modifying the technique for spatio-temporal contexts may yield further insights into the changing dynamics of firearm violence. In the future, we will transfer the knowledge of county-specific variable selection for gun-violence data and then describe the perspective of transfer learning in county-specific variable selection, which eliminates the need for repeated computation.
 \section*{Data and Code Availability}
 The data is open source. The code is available in \url{https://github.com/debjoythakur/-Local-Variable-and-Neighborhood-Selection-for-Firearm-Fatality-in-South-East-USA}.
 \section*{Funding}
     Bandyopadhyay's research was funded by National Science Foundation grants CMMI-2210840 and DMS-2514857.
\printbibliography
\newpage
\section*{Supplementary Material}
\subsection*{Motivation}
 \newpage
 \subsection*{Proof of Theorem~\ref{thm:sparsity}}
\begin{proof}
    Let's consider $\alpha_n = \sqrt{K_n / N_n}$ we require to prove for a given $\epsilon > 0$ there exists a large constant $C$ such that
\[
\mathbb P\!\left(\sup_{\|\bm u\|=C} \mathcal L_n^{(i)}(\bm\eta^0 + \alpha_n \bm u) < \mathcal L_n^{(i)}(\bm\eta^0)\right) \ge 1-\epsilon.
\]
Now define the objective function
\begin{equation}\label{eq:vnu}
    \begin{split}
        V_n(\bm u)
&= \mathcal L_n^{(i)}(\bm\eta^0 + \alpha_n \bm u)
   - \mathcal L_n^{(i)}(\bm\eta^0) \nonumber\\
&= \frac{1}{N_n}
   \Bigl\{ L^{(i)}(\bm\eta^0 + \alpha_n \bm u)
          - L^{(i)}(\bm\eta^0) \Bigr\} \nonumber\\
&\quad
 - \sum_{k=1}^{P_0}
 \Bigl\{ \mathcal{P}_{\lambda_n}(|\eta_k^0 + \alpha_n u_k|)
        - \mathcal{P}_{\lambda_n}(|\eta_k^0|) \Bigr\}.
    \end{split}
\end{equation}
Using a second-order Taylor expansion, there exists
$t\in(0,1)$ such that $\bm\eta^* = \bm\eta^0 + t\alpha_n \bm u$ then the objective function in (\ref{eq:vnu}) becomes:
\begin{align*}
V_n(\bm u)
&= \frac{\alpha_n}{N_n}
   \bm\omega_{0n}^{(i)}(\bm\eta^0)^{\top}\bm u
 - \frac{1}{2N_n}\alpha_n^2
   \bm u^{\top}\bm\Omega_{00n}^{(i)}(\bm\eta^*)\bm u \nonumber\\
&\quad
 - \sum_{k=1}^{P_0}
 \Bigl\{ \mathcal{P}_{\lambda_n}(|\eta_k^0 + \alpha_n u_k|)
        - \mathcal{P}_{\lambda_n}(|\eta_k^0|) \Bigr\}\\
    &= \frac{\alpha_n}{N_n}
   \bm\omega_{0n}^{(i)}(\bm\eta^0)^{\top}\bm u
 - \frac{1}{2N_n}\alpha_n^2
   \bm u^{\top}\bm\Omega_{00n}^{(i)}(\bm\eta^0)\bm u \nonumber\\
&\quad
 + \frac{1}{2N_n}\alpha_n^2
   \bm u^{\top}
   \Bigl\{
   \bm\Omega_{00n}^{(i)}(\bm\eta^0)
 - \bm\Omega_{00n}^{(i)}(\bm\eta^*)
   \Bigr\}\bm u \nonumber\\
&\quad
 - \sum_{k=1}^{P_0}
 \Bigl\{ \mathcal{P}_{\lambda_n}(|\eta_k^0 + \alpha_n u_k|)
        - \mathcal{P}_{\lambda_n}(|\eta_k^0|) \Bigr\}\\
& = T_1 + T_2 + T_3 + T_4.
\end{align*}
Now using A-\ref{ass:score} we can derive,
\begin{equation}\label{eq:upperbound T1}
    |T_1|
\le
\alpha_n
\left\|
\frac{1}{N_n}\bm\omega_{0n}^{(i)}(\bm\eta^0)
\right\|
\|\bm u\|
=
O_p(\alpha_n^2)\|\bm u\|.
\end{equation}
Next using A-\ref{ass:fisher} we deduce,
\begin{equation}\label{eq:upperbound T2}
    T_2
=
-\frac{\alpha_n^2}{2}
\bm u^{\top}
\left(
\frac{1}{N_n}\bm\Omega_{00n}^{(i)}(\bm\eta^0)
\right)
\bm u
\le
-\frac{\alpha_n^2}{2}\Lambda_n\|\bm u\|^2.
\end{equation}
There exists $\tilde{\bm\eta}$ on the line segment joining
$\bm\eta^0$ and $\bm\eta^*$ such that
\begin{align}\label{eq:T3_simplify}
T_3&= \frac{1}{2N_n}\alpha_n^2
   \bm u^{\top}
   \Bigl\{
   \bm\Omega_{00n}^{(i)}(\bm\eta^0)
 - \bm\Omega_{00n}^{(i)}(\bm\eta^*)
   \Bigr\}\bm u\\
&=
\frac{1}{2N_n}\alpha_n^2
\sum_{j=1}^{N_n}
\frac{\exp(\bm z_j^{\top}\tilde{\bm\eta})}
     {(1+\exp(\bm z_j^{\top}\tilde{\bm\eta}))^2}
 \bm z_j^{\top}(\bm\eta^0-\bm\eta^*)
 (\bm z_j^{\top}\bm u)^2 \nonumber\\
&=
- t\frac{\alpha_n^3}{2N_n}
\sum_{j=1}^{N_n}
\frac{\exp(\bm z_j^{\top}\tilde{\bm\eta})}
     {(1+\exp(\bm z_j^{\top}\tilde{\bm\eta}))^2}
 (\bm z_j^{\top}\bm u)^3\nonumber.
\end{align}
Using A-\ref{ass:bounded-z} from \eqref{eq:T3_simplify} we derive that,
\begin{equation}\label{eq:bounded T3}
    |T_3|
\le
\frac{1}{2N_n}\alpha_n^3
\sum_{j=1}^{N_n}
\tilde C
\|\bm z_j\|^3
\|\bm u\|^3
=O(\alpha_n^3)\|\bm u\|^3.
\end{equation}
Using Taylor's second-order expansion for the SCAD penalty term
\[
\begin{aligned}
    T_4
&=
-\sum_{k=1}^{P_0}
\Bigl\{
\mathcal{P}_{\lambda_n}(|\eta_k^0+\alpha_n u_k|)
-
\mathcal{P}_{\lambda_n}(|\eta_k^0|)
\Bigr\}\\
&=
-\sum_{k=1}^{P_0}
\alpha_n
\mathcal{P}'_{\lambda_n}(|\eta_k^0|)
\operatorname{sgn}(\eta_k^0)u_k
-
\alpha_n^2
\sum_{k=1}^{P_0}
\mathcal{P}''_{\lambda_n}(|\eta_k^0|)
u_k^2(1+o(1)).
\end{aligned}
\]
Using A-\ref{ass:penalty-parameter} and A-\ref{ass:penalty}, we derive the upper bound of the penalty term as follows:
\begin{equation}\label{eq:bounded T4}
    |T_4|
\le
\sqrt{P_0}\,\alpha_n a_n \|\bm u\|
+
\alpha_n^2
\max_{1\le k\le P_0}
\{|\mathcal{P}^{\prime\prime}_{\lambda_n}(|\eta_k^0|)|:\eta_k^0\neq0\}
\|\bm u\|^2.
\end{equation}
For enough large $N_n$, fixing $\|\bm u\|=C$,
combining (\ref{eq:upperbound T1}), (\ref{eq:upperbound T2}), (\ref{eq:bounded T3}), (\ref{eq:bounded T4}) and A-\ref{ass:penalty}, we get
\begin{equation}\label{eq:bound-vnu}
    \begin{aligned}
        V_n(\bm u) &= T_2 + (T_1+ T_3 + T_4)\le T_2 +(|T_1| +  |T_3| + |T_4|)\\
&\le
O_p(\alpha_n^2)C
-
\frac{\Lambda_n}{2}\alpha_n^2 C^2
+O(\alpha_n^3) C^3
+
\sqrt{P_0}\alpha_n a_n C + \alpha_n^2 b_n C^2.
    \end{aligned}
\end{equation}
In \eqref{eq:bound-vnu} the term $T_1$ is dominated by $\frac{\Lambda_n}{2}\alpha_n^2 C^2$ for fixed positive $C$. Since $\alpha_n \to 0$ as $N_n \to \infty$ therefore for a fixed $C$ the term $\alpha_n^3 = o(\alpha_n^2)$ raised from $T_3$ in \eqref{eq:bound-vnu} therefore the third term is also bounded by $\frac{\Lambda_n}{2}\alpha_n^2 C^2$. According to the A-\ref{ass:penalty} the terms $\alpha_n a_n C = O(\alpha_n^2 C^2)$ and $\alpha_n^2 b_n C^2 = o(\Lambda_n \alpha_n^2 C^2)$. Hence for a given $\epsilon>0$, we can pick $C>0$ such that
\[
\mathbb P\!\left(
\sup_{\|\bm u\|=C} V_n(\bm u) < 0
\right)
\ge 1-\varepsilon.
\]

Let $\mathcal{A}_n^{(i)}=\{1,\dots,P_0\}$ be the index set of nonzero components of
$\bm\eta^0$ and $\mathcal{B}_n^{(i)}=\{P_0+1,\dots,K_n\}$ its complement. Recall the
penalized objective function
\[
\mathcal{L}^{(i)}_n(\bm\eta)
=
\frac{1}{N_n}L^{(i)}(\bm\eta)
-\sum_{k=1}^{K_n}\mathcal{P}_{\lambda_n}(|\eta_k|).
\]
From the previous part, we derive that there exists a local maximizer
$\widehat{\bm\eta}$ such that
$\|\widehat{\bm\eta}-\bm\eta^0\|=O_p(\alpha_n)$ with probability tending to one. Next we need to establish that
\[
\mathbb P\Bigl(\widehat\eta^{(i)}_k=0\ \text{for all }k\in\mathcal{B}_n^{(i)}\Bigr)\to 1.
\]

Fix any $k\in\mathcal{B}_n^{(i)}$, so that $\eta_k^0=0$. We prove that there exists
$\varepsilon_n>0$ such that for all $\bm\eta$ satisfying
$\|\bm\eta-\bm\eta^0\|=O_p(\alpha_n)$,
\begin{equation}\label{eq:A3A4-our}
\begin{aligned}
\frac{\partial \mathcal{L}^{(i)}_n(\bm\eta)}{\partial \eta_k} &< 0,
\qquad 0<\eta_k<\varepsilon_n,\\
\frac{\partial \mathcal{L}^{(i)}_n(\bm\eta)}{\partial \eta_k} &> 0,
\qquad -\varepsilon_n<\eta_k<0,
\end{aligned}
\end{equation}
with probability tending to one, which implies $\widehat\eta_k=0$. This implies that $\mathcal L_n^{(i)}$ is decreasing for $\eta_k>0$ and increasing for $\eta_k<0$ in a neighborhood of $0$,
hence any local maximizer must satisfy $\widehat\eta_k=0$. For $|\eta_k|\le \varepsilon_n$, by Taylor expansion of the score around $\bm\eta^0$, there exists $\bm\eta^\star$ on the line segment between $\bm\eta$ and $\bm\eta^0$ such that
\begin{align}
\frac{\partial \mathcal L_n^{(i)}(\bm\eta)}{\partial \eta_k}
&=
\underbrace{\frac{1}{N_n}\frac{\partial L^{(i)}(\bm\eta^0)}{\partial \eta_k}}_{T_5}
+
\underbrace{\frac{1}{N_n}\sum_{l=1}^{K_n}
\frac{\partial^2 L^{(i)}(\bm\eta^0)}{\partial\eta_k\partial\eta_l}(\eta_l-\eta_l^0)}_{T_6}
\notag\\
&\quad+
\underbrace{\frac{1}{N_n}\sum_{l=1}^{K_n}
\Bigg[
\frac{\partial^2 L^{(i)}(\bm\eta^\star)}{\partial\eta_k\partial\eta_l}
-\frac{\partial^2 L^{(i)}(\bm\eta^0)}{\partial\eta_k\partial\eta_l}
\Bigg](\eta_l-\eta_l^0)}_{T_7}
-\underbrace{\bigl(\mathcal{P}'_{\lambda_n}(|\eta_k|)\operatorname{sgn}(\eta_k)\bigr)}_{T_8}.
\label{eq:T5T8}
\end{align}
Let's consider the $\alpha_n$-radius ball around true $\bm \eta_0$ as $\mathcal E_n:=\{\|\bm\eta-\bm\eta^0\|\le \alpha_n\}$. By utilizing A-\ref{ass:score} we can derive,
\begin{equation}\label{eq:T5-A3}
|T_5|
=
\left|\frac{1}{N_n}\frac{\partial L^{(i)}(\bm\eta^0)}{\partial \eta_k}\right|
\le
\left\|\frac{1}{N_n}\bm\omega_n^{(i)}(\bm\eta^0)\right\|
=
O_p(\alpha_n).
\end{equation}

Recall
\[
T_6
=
\frac{1}{N_n}\sum_{l=1}^{K_n}
\frac{\partial^2 L^{(i)}(\bm\eta^0)}{\partial\eta_k\partial\eta_l}
(\eta_l-\eta_l^0)= -\frac{1}{N_n}\sum_{j=1}^{N_n}\sum_{l=1}^{K_n} \frac{\tilde{\mu}_j^0}{1+\sigma \tilde{\mu}_j^0}\, z_{jk}z_{jl}(\eta_l-\eta_l^0),
\]
 Thus using A-\ref{ass:bounded-z} we derive
\begin{align}
|T_6|
&\le
\sum_{l=1}^{K_n} C_\mu C_z^2\,|\eta_l-\eta_l^0|
=
C_\mu C_z^2 \|\bm\eta-\bm\eta^0\|_1 \notag\\
&\le
C_\mu C_z^2 \sqrt{K_n}\,\|\bm\eta-\bm\eta^0\|
=
O_P\!\left(\sqrt{K_n}\alpha_n\right)
\quad\text{on }\mathcal E_n.
\label{eq:T6-clean}
\end{align}
Recall from \eqref{eq:T5T8} we have 
\[
T_7
=
\frac{1}{N_n}\sum_{l=1}^{K_n}
\Bigg[
\frac{\partial^2 L^{(i)}(\bm\eta^\star)}{\partial\eta_k\partial\eta_l}
-
\frac{\partial^2 L^{(i)}(\bm\eta^0)}{\partial\eta_k\partial\eta_l}
\Bigg](\eta_l-\eta_l^0),
\]
where $\bm\eta^\star$ lies on the segment joining $\bm\eta$ and $\bm\eta^0$.
By a first-order Taylor expansion of the Hessian, we have,
\[
\frac{\partial^2 L^{(i)}(\bm\eta^\star)}{\partial\eta_k\partial\eta_l}
-
\frac{\partial^2 L^{(i)}(\bm\eta^0)}{\partial\eta_k\partial\eta_l}
=
\sum_{m=1}^{K_n}
\frac{\partial^3 L^{(i)}(\tilde{\bm\eta})}
{\partial\eta_k\partial\eta_l\partial\eta_m}
(\eta_m-\eta_m^0)
\]
for some $\tilde{\bm\eta}$ between $\bm\eta^\star$ and $\bm\eta^0$.
Under the assumption A-\ref{ass:thirdderiv} we finally obtain:
\begin{align}
|T_7|
&\le
\sum_{l=1}^{K_n}\sum_{m=1}^{K_n}
C_3\,|\eta_m-\eta_m^0|\,|\eta_l-\eta_l^0|
\notag\\
&=
C_3 \|\bm\eta-\bm\eta^0\|_1^2
\le
C_3 K_n \|\bm\eta-\bm\eta^0\|^2
=
O_p(K_n\alpha_n^2)
\quad\text{on }\mathcal E_n.
\label{eq:T7-clean}
\end{align}
Combining \eqref{eq:T5-A3}, \eqref{eq:T6-clean}, and \eqref{eq:T7-clean}, we obtain
\[
\sup_{\eta_k \in \mathcal E_n}|T_5|+|T_6|+|T_7|
=
O_p(\alpha_n)
+
O_p(\sqrt{K_n}\alpha_n)
+
O_p(K_n\alpha_n^2).
\]
\medskip
Let $0<\eta_k<\varepsilon_n$, From
\eqref{eq:T5T8},
\[
\begin{aligned}
    \frac{\partial \mathcal L_n^{(i)}(\bm\eta)}{\partial \eta_k}
=
T_5+T_6+T_7-\mathcal P'_{\lambda_n}(|\eta_k|)\\
\implies \frac{\partial \mathcal L_n^{(i)}(\bm\eta)}{\partial \eta_k}
\le
|T_5|+|T_6|+|T_7|
-\mathcal P'_{\lambda_n}(|\eta_k|).
\end{aligned}
\]
Using the conditions A-\ref{ass:penalty-parameter} and (\ref{ass:sparsity-penalty}) we can say for any $\gamma \in (0, \kappa_0)$ there exists $N_0, \varepsilon_n >0$ such that $\inf_{0< t < \varepsilon_n} \mathcal{P}_{\lambda_n}^\prime(t) \ge (\kappa_0 - \gamma)\lambda_n$ for all $N_n > N_0$. Now setting $\gamma = \frac{\kappa_0}{2}$ there exists $N_0, \varepsilon_n >0$ such that $\mathcal{P}_{\lambda_n}^\prime(t) \ge \frac{\kappa_0}{2}\lambda_n$ for all $t\in (0, \varepsilon_n)$ and for all $N_n \ge N_0$. Thus, we finally have
\[
\frac{\partial \mathcal L_n^{(i)}(\bm\eta)}{\partial \eta_k}
\le
-\frac{\kappa_0}{2}\lambda_n\{1+o_P(1)\}<0
\quad\text{w.p.a.1}.
\]
This proves the consistency of sparsity selection. A similar approach will be carried out for $-\varepsilon_n<\eta_k<0$.
\end{proof}
\newpage
\subsection*{Additional Simulation Results}
\newpage
\subsection*{Additional Real Data Analysis Results}

\end{document}